\RequirePackage{ifpdf}

\documentclass[a4paper,11pt]{article}
\pdfoutput=1 

\usepackage{jheppub} 

\usepackage[T1]{fontenc} 

\usepackage{ifpdf} 

\usepackage{flexisym}
\usepackage{breqn}
\usepackage{comment}

\usepackage{epsf,epsfig,epstopdf}
\usepackage{graphics}
\usepackage{graphicx}
\usepackage[usenames,dvipsnames]{xcolor}
\definecolor{darkgreen}{RGB}{34,139,34}

\usepackage{enumerate}

\usepackage{amsmath,amsfonts,amssymb}

\author{Costas G. Papadopoulos$^{a,b}$, Damiano Tommasini$^{a}$ and Christopher Wever$^{a,c}$}

\preprint{TTP15-042}

\affiliation{$^{a}$Institute of Nuclear and Particle Physics, NCSR `Demokritos', Agia Paraskevi, 15310, Greece}
\affiliation{$^{b}$MTA-DE Particle Physics Group, University of Debrecen, H-4010 Debrecen, Hungary}
\affiliation{$^{c}$Institute for Theoretical Particle Physics (TTP), Karlsruhe Institute of Technology, Engesserstra{\ss}e 7, D-76128 Karlsruhe, Germany $\&$ 
Institute for Nuclear Physics (IKP), Karlsruhe Institute of Technology, Hermann-von-Helmholtz-Platz 1, D-76344 Eggenstein-Leopoldshafen, Germany}

\emailAdd{costas.papadopoulos@cern.ch, tommasini@inp.demokritos.gr, wever@inp.demokritos.gr}


\title{The Pentabox Master Integrals with the Simplified Differential Equations approach}

\abstract{
We present the calculation of massless two-loop Master Integrals relevant to five-point amplitudes with one off-shell external leg and derive the complete set 
of planar Master Integrals with five on-mass-shell legs, that contribute 
to many $2\to 3$ amplitudes of interest  at the LHC, as for instance 
 three jet production, $\gamma, V, H +2$ jets etc., based on the  Simplified Differential 
Equations approach.
}

\ifpdf
\fi

\begin{document}
\unitlength1cm
\maketitle
\allowdisplaybreaks

\section{Introduction}

With LHC delivering collisions at the highest energy achieved so far, 13 TeV, experiments are analysing data corresponding to an integrated luminosity of $42$ pb$^{-1}$~\cite{Khachatryan:2015uqb} and $85$ pb$^{-1}$~\cite{Atlas}, 
as well as those already collected at an energy of 8 TeV and an integrated luminosity of $20.3$ fb$^{-1}$~\cite{Aad:2015nda} and $19.7$ fb$^{-1}$~\cite{Khachatryan:2015tzo}. 
In order to keep up with the increasing experimental accuracy as more data is collected, more precise theoretical predictions and higher loop calculations are required~\cite{Andersen:2014efa}.

In the last ten years our understanding of the reduction of one-loop amplitudes to a set of Master Integrals ({\bf MI}) based on unitarity methods~\cite{Bern:1994cg,Bern:1994zx} and at the integrand level via the OPP 
method~\cite{Ossola:2006us,Ossola:2008xq}, has drastically changed the way one-loop calculations are preformed leading to many fully automated numerical tools (some reviews on the topic are~\cite{AlcarazMaestre:2012vp,Ellis:2011cr}).
In the recent years, a lot of progress has been made also towards the extension of these reduction methods for two-loop amplitudes at the integral~\cite{Gluza:2010ws,Kosower:2011ty,CaronHuot:2012ab} 
as well as the integrand~\cite{Mastrolia:2011pr,Badger:2012dp,Badger:2013gxa,Papadopoulos:2013hra} level. Contrary to the one-loop case, where MI have been known for a long time already~\cite{'tHooft:1978xw}, a complete library of MI at two-loops is still missing.
At the moment this is the main obstacle to obtain a fully automated NNLO calculation framework similar to the one-loop one, that will satisfy the anticipated precision requirements at the LHC~\cite{Butterworth:2014efa}.

Following the work of~\cite{Goncharov:1998kja,Remiddi:1999ew,Goncharov:2001iea}, there has been a building consensus that the so-called {\it Goncharov Polylogarithms} ({\bf GPs}) form a functional basis for many MI. A very successful method for calculating MI and expressing them in terms of GPs is the differential equations ({\bf DE}) approach~\cite{Kotikov:1990kg,Kotikov:1991pm,Bern:1992em,Remiddi:1997ny,Gehrmann:1999as,Henn:2013pwa}, which has been used in the past two decades to calculate various MI at two-loops ~\cite{Caffo:1998du,Gehrmann:1999as,Gehrmann:2000zt,Gehrmann:2001ck,Bonciani:2003te,Laporta:2004rb,Bonciani:2008az,Gehrmann:2013cxs,vonManteuffel:2013uoa,Henn:2014lfa,Gehrmann:2014bfa,Caola:2014lpa,Papadopoulos:2014hla,Gehrmann:2015ora}. 
In~\cite{Papadopoulos:2014lla} a variant of the traditional DE approach to MI was presented, which was coined the Simplified Differential Equations ({\bf SDE}) approach. 
In this paper we present a further application of this method, concerning the calculation of planar massless MI relevant to five-point amplitudes with one off-shell leg, as well as the complete set of planar MI for five-point on-shell amplitudes.
This is an important step towards the calculation of the full set of MI with up to eight internal propagators needed to realise a fully automated reduction scheme, \`a la OPP, for NNLO QCD. 
 
Pentabox integrals are needed in particular in order to compute NNLO QCD corrections to several processes of interest at LHC~\cite{Andersen:2014efa}. The $pp\rightarrow H+2$jets can be used to measure the $HWW$ coupling to a $5\%$ accuracy 
with 300 fb$^{-1}$ data. The $pp\rightarrow 3$jets to study the ratio of $3-$jet to $2-$jet cross sections and measure the running of the strong coupling constant. 
The   $pp\rightarrow V+2$jets for PDF determination and background studies for multi-jet final states.

The paper is organized as follows. In Section 2 we set the parameterization and notation of the variables describing the two-loop MI of interest. 
In Section 3 we discuss the DE obtained, and the results for the pentabox MI. 
We conclude in Section 4 and provide an overview of the topic and some perspective for future developments. 
In the Appendix \ref{x=1} we present details on the derivation of the planar pentabox MI with on-shell legs and in the Appendix \ref{expbyreg} we give a few characteristic examples on how the boundary conditions are properly reproduced in our approach by the DE. 
Finally in the ancillary files~\cite{results}, we provide our analytic results for all  two-loop MI in terms of Goncharov polylogarithms together with explicit numerical results.

\section{The pentabox integrals}

The MI in this paper will be calculated with the SDE approach~\cite{Papadopoulos:2014lla}.
Assume that one is interested in calculating an $l-$loop Feynman integral with external momenta $\{p_j\}$, considered incoming, and internal propagators that are massless.
Any $l-$loop  Feynman integral can be then written as
\begin{equation}
G_{a_1\cdots a_n}(\{p_j\},\epsilon)=\int\left(\prod_{r=1}^l \frac{d^dk_r}{i\pi^{d/2}}\right)\frac{1}{D_1^{a_1}\cdots D_n^{a_n}}, \hspace{0.5 cm}
D_i=\left(c_{ij}k_j+d_{ij}p_j\right)^2,\,\, d=4-2\epsilon
\label{eq:loopgen}
\end{equation}
with matrices $\{c_{ij}\}$ and $\{d_{ij}\}$ determined by the topology and the momentum flow of the graph, and
the denominators are defined in such a way that all scalar product invariants can be written as a linear combination of them. The exponents $a_i$ are integers and may be negative in order to accommodate irreducible numerators.

Any integral $G_{a_1\cdots a_n}$ may be written as a linear combination of a finite subset of such integrals, called Master Integrals,  
with coefficients depending on the independent scalar products, $s_{ij}=p_i\cdot p_j$, and space-time dimension $d$, by the use of {\it integration by parts} ({\bf IBP}) identities~\cite{Chetyrkin:1981qh,Tkachov:1981wb,Laporta:2001dd}.  
In the traditional DE method, the Master Integrals
are differentiated with respect to $p_i \cdot \frac{\partial}{\partial p_j}$ 
and the resulting integrals are reduced by IBP to give a linear system of first order DE~\cite{Kotikov:1990kg,Remiddi:1997ny}.  
 The  invariants, $s_{ij}$, are then parametrised in terms of dimensionless variables, defined on a case by case basis, so that the resulting DE can be solved in terms of GPs. 
Usually boundary terms corresponding to the appropriate limits of the chosen parameters have to be calculated 
using for instance expansion by regions techniques~\cite{Beneke:1997zp,Smirnov:2002pj}.

SDE approach~\cite{Papadopoulos:2014lla} is an attempt not only to simplify, but also to systematize, as much as possible, the derivation of the appropriate system of DE satisfied by the MI. 
To this end 
the external incoming momenta are {\it parametrized} linearly in terms of $x$ as $p_i(x)=p_i+(1-x)q_i$, where the $q_i$'s are a linear combination of the  momenta $\{p_i\}$ such that $\sum_iq_i=0$. If $p_i^2=0$, the parameter $x$ captures the off-shell-ness of the external leg. 
The class of Feynman integrals in (\ref{eq:loopgen}) are now dependent on $x$ through the external momenta:
\begin{equation}
G_{a_1\cdots a_n}(\{s_{ij}\},\epsilon;x)=\int\left(\prod_{r=1}^l \frac{d^dk_r}{i\pi^{d/2}}\right)\frac{1}{D_1^{a_1}\cdots D_n^{a_n}}, 
\;\;\;
D_i=\left( c_{ij}k_j+d_{ij}p_j(x) \right)^2.
\label{eq:loopgenx}
\end{equation}

By introducing the dimensionless parameter $x$, the array of MI, $\mathbf{G}(\{s_{ij}\},\epsilon;x)$, which now depends on $x$, satisfies
\begin{equation}
\frac{\partial}{\partial x} \mathbf{G}(\{s_{ij}\},\epsilon;x)=\mathbf{H}(\{ s_{ij}\},\epsilon;x)\mathbf{G}(\{s_{ij}\},\epsilon;x)\label{eq:DEx}
\end{equation}
a system of differential equations in one independent variable, where $\mathbf{H}$ is a matrix whose elements are rational functions of the kinematics $\{s_{ij}\equiv p_i\cdot p_j\}$, of $x$ and of $\epsilon$. 

Experience up to now shows that this simple parametrization can be used universally to deal with up to six kinematical scales involved, as is the case we will present in this paper.
 The expected benefit is that the integration of the DE naturally captures the expressibility of MI in terms of GPs and more importantly makes the problem 
 {\it independent of the number of kinematical scales} (independent invariants) involved. 
Note that as $x\rightarrow 1$, the original configuration of the loop integrals (\ref{eq:loopgen}) is reproduced, which corresponds to a simpler one with one scale less.

\begin{figure}[t!]
\centering
\includegraphics[width=0.20 \linewidth]{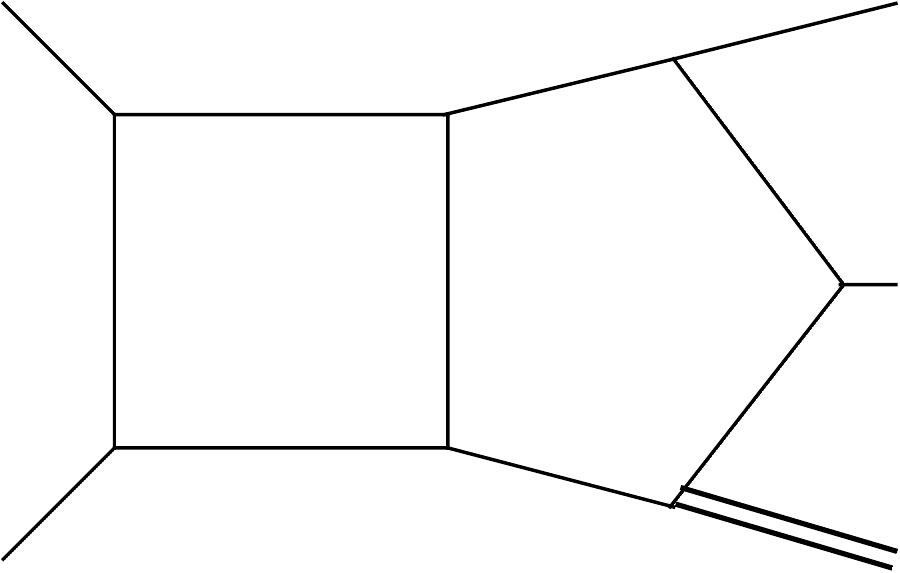} \hspace{0.4 cm}
\includegraphics[width=0.20 \linewidth]{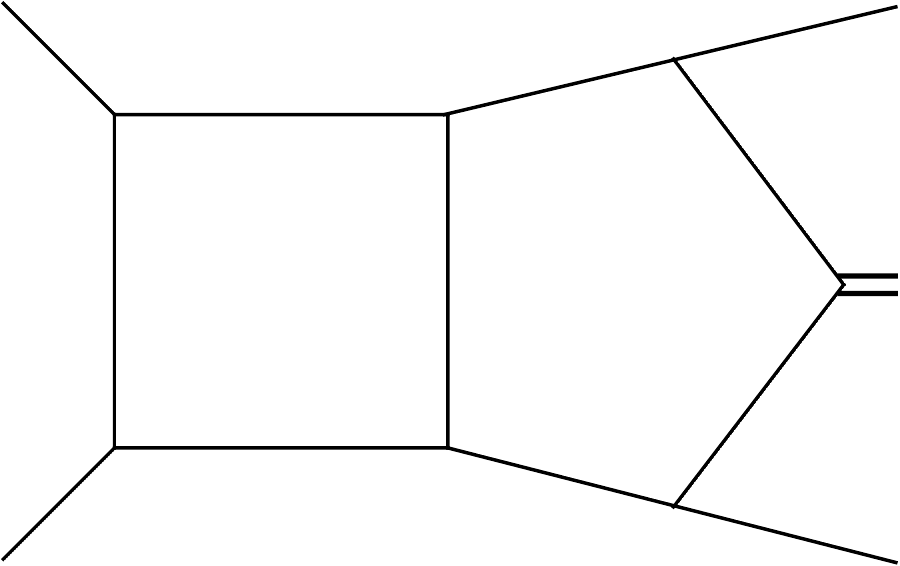} \hspace{0.4 cm}
\includegraphics[width=0.20 \linewidth]{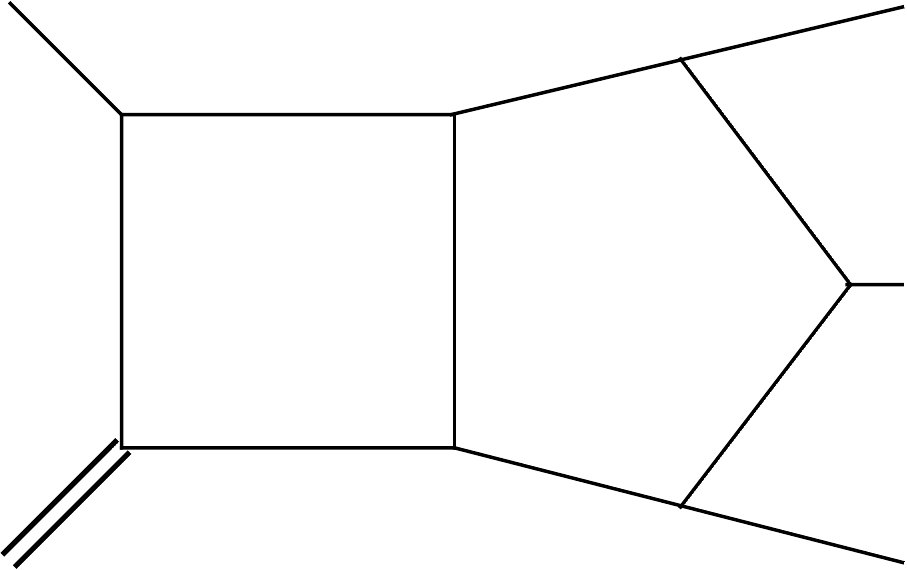}
  \caption{The three planar pentaboxes of the families $P_1$ (left), $P_2$ (middle) and $P_3$ (right) with one external massive leg.}
  \label{fig:param-P}
\end{figure}

\begin{figure}[t!]
\centering
\includegraphics[width=0.20 \linewidth]{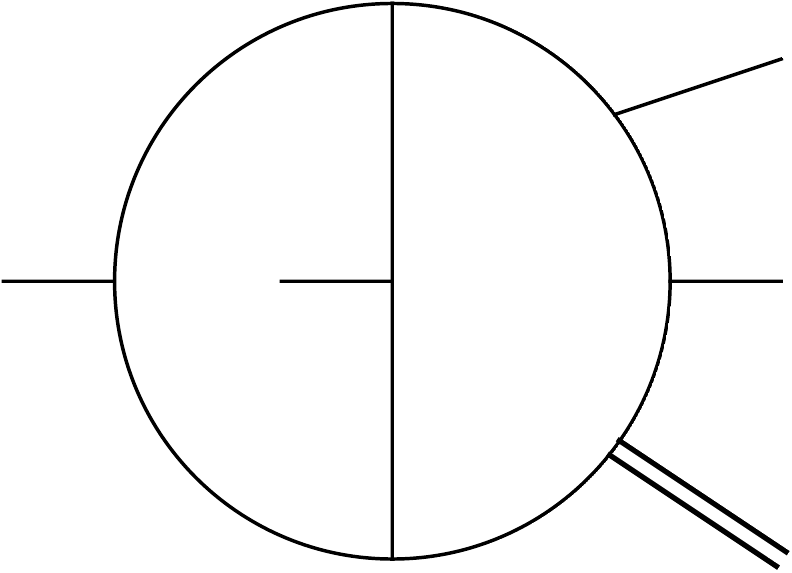} \hspace{0.6 cm}
\includegraphics[width=0.20 \linewidth]{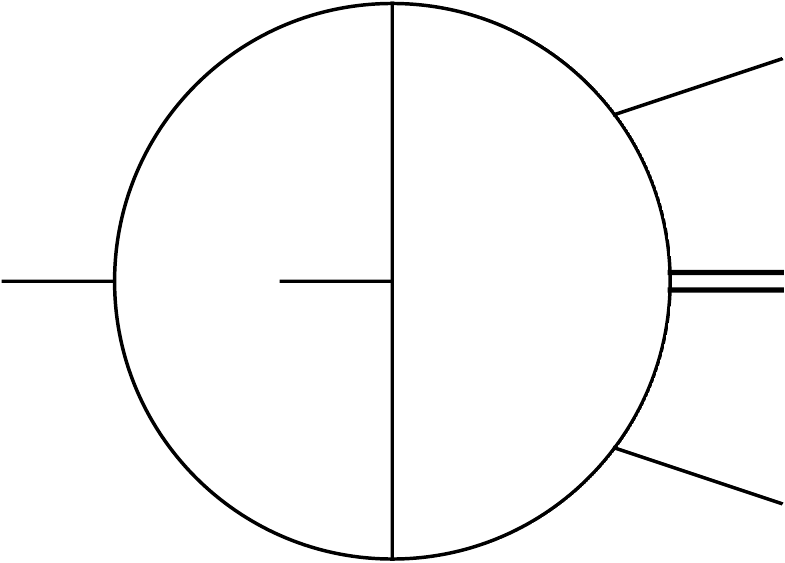} \hspace{0.6 cm}
\includegraphics[width=0.20 \linewidth]{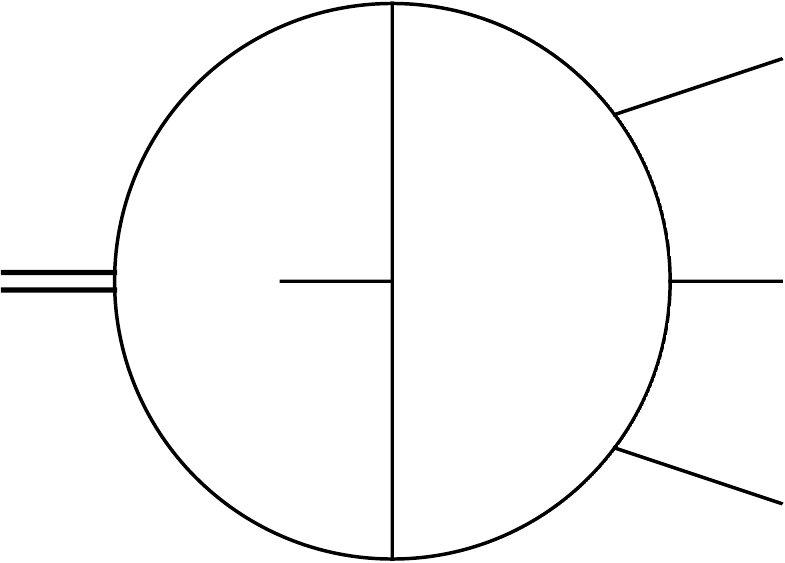} \\[12pt]
\includegraphics[width=0.20 \linewidth]{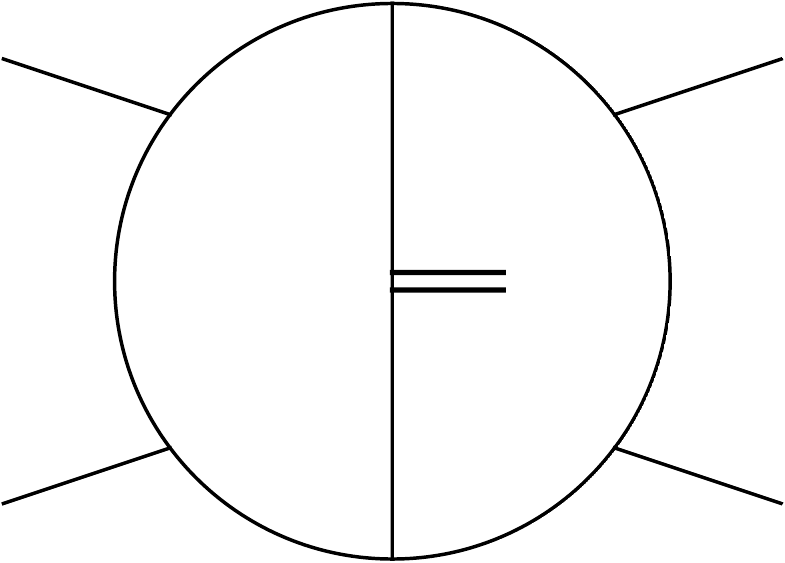} \hspace{0.6 cm}
\includegraphics[width=0.20 \linewidth]{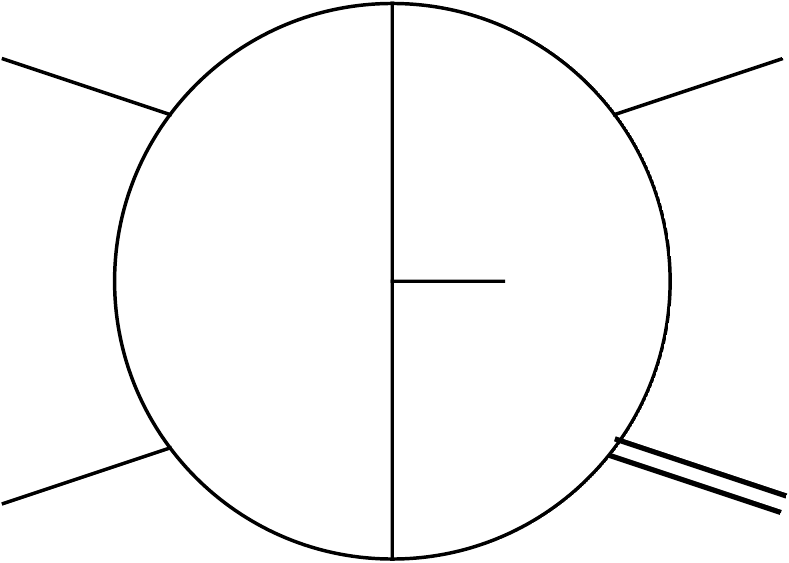}
  \caption{The five non-planar families with one external massive leg.}
  \label{fig:param-N}
\end{figure}

More specifically we are interested in calculating the MI of two-loop QCD five-point amplitudes. As it is an inherent characteristic of the
SDE method to interpolate among different kinematical configurations of the external momenta the starting point is to compute
five-point amplitudes with one off-shell leg. These amplitudes contribute to the production i.e., of one massive final state $V$, plus two massless final states $j_1,j_2$ at the LHC:
\begin{gather}
p(q_1)p'(q_2)\rightarrow V(q_3)j_1(q_4)j_2(q_5), \ \ q_1^2=q_2^2=0, \ \ q_3^2=M_3^2, \ \ q_4^2=q_5^2=0.\label{eq:ppVV}
\end{gather}
The colliding partons have massless momenta $q_1,q_2$, while the outgoing massive and the two massless particles have momenta $q_3$ and $q_4,q_5$ respectively. Of course, by appropriately taking the limit
$x=1$ the pentabox MI with all external massless momenta on-shell will be obtained, that are relevant for instance to the three-jet production
\begin{gather}
p(q_1)p'(q_2)\rightarrow j_1(q_3)j_2(q_4)j_3(q_5), \ \ q_i^2=0.\label{eq:pp3j}
\end{gather}

For the off-shell case $M_3^2\neq 0$, there are in total three families of planar MI whose members with the maximum amount of denominators, namely eight, are graphically shown in Figure \ref{fig:param-P}. Similarly, there are five non-planar families of MI as given in Figure \ref{fig:param-N}. We have checked that the other five-point integrals with one massive external leg are reducible to MI in one of these eight MI families. The two-loop planar (Fig. \ref{fig:param-P}) and non-planar (Fig. \ref{fig:param-N}) diagrams contributing to (\ref{eq:ppVV}) have not been calculated yet. In fact by taking the limit $x=1$ all planar graphs for massless on-shell external momenta are derived as well~\footnote{During the writing of the present paper, some results related to massless 
planar pentaboxes appeared in~\cite{Gehrmann:2015bfy}.}. 
In this paper we calculate all MI in the family $P_1$ as well as all the the on-shell MI as $M_3^2\rightarrow 0$. We use the {\tt c++} implementation of the program {\bf FIRE}~\cite{Smirnov:2014hma} to perform the IBP reduction to the set of MI in $P_1$.

\begin{figure}[t!]
\centering
\includegraphics[width=0.29 \linewidth]{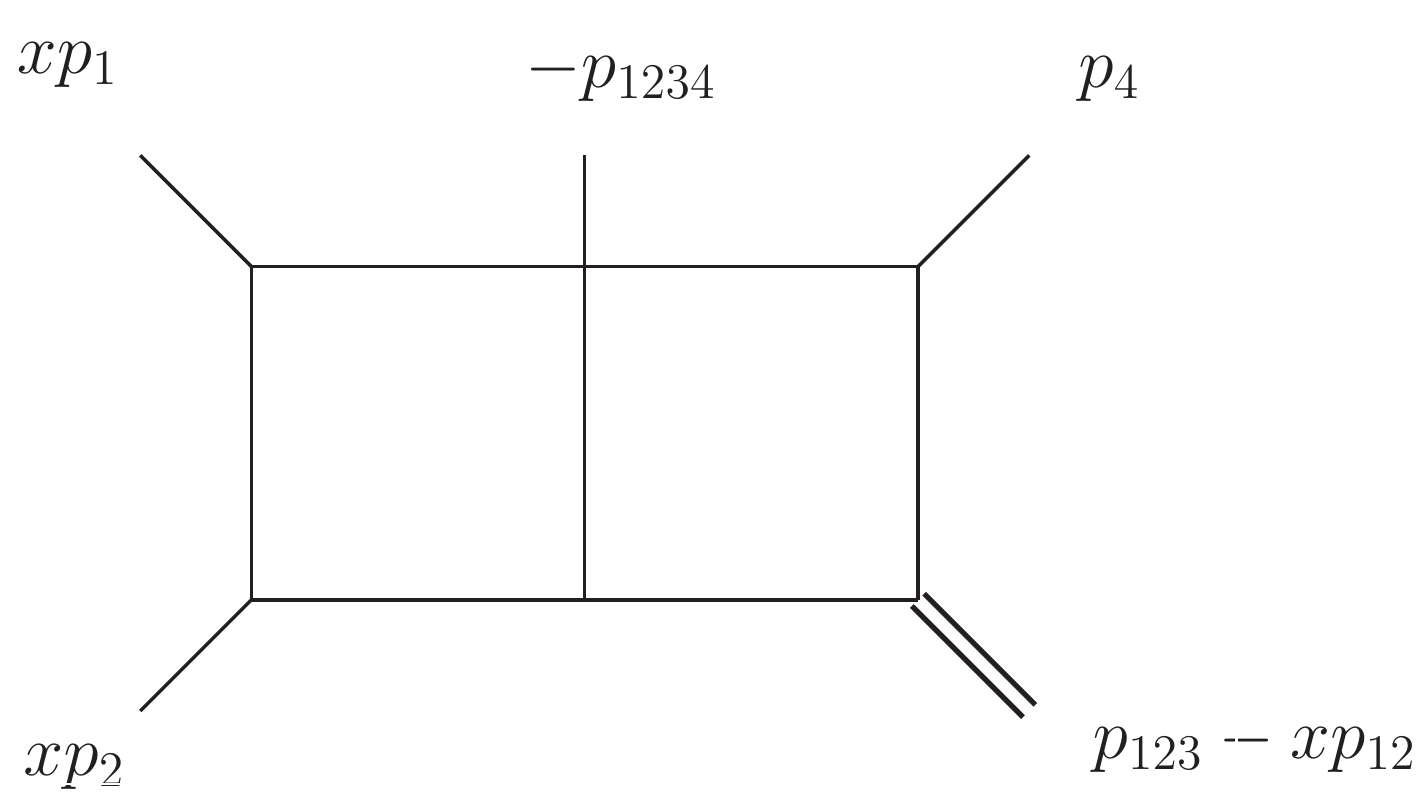} \hspace{0.4 cm}
\includegraphics[width=0.29 \linewidth]{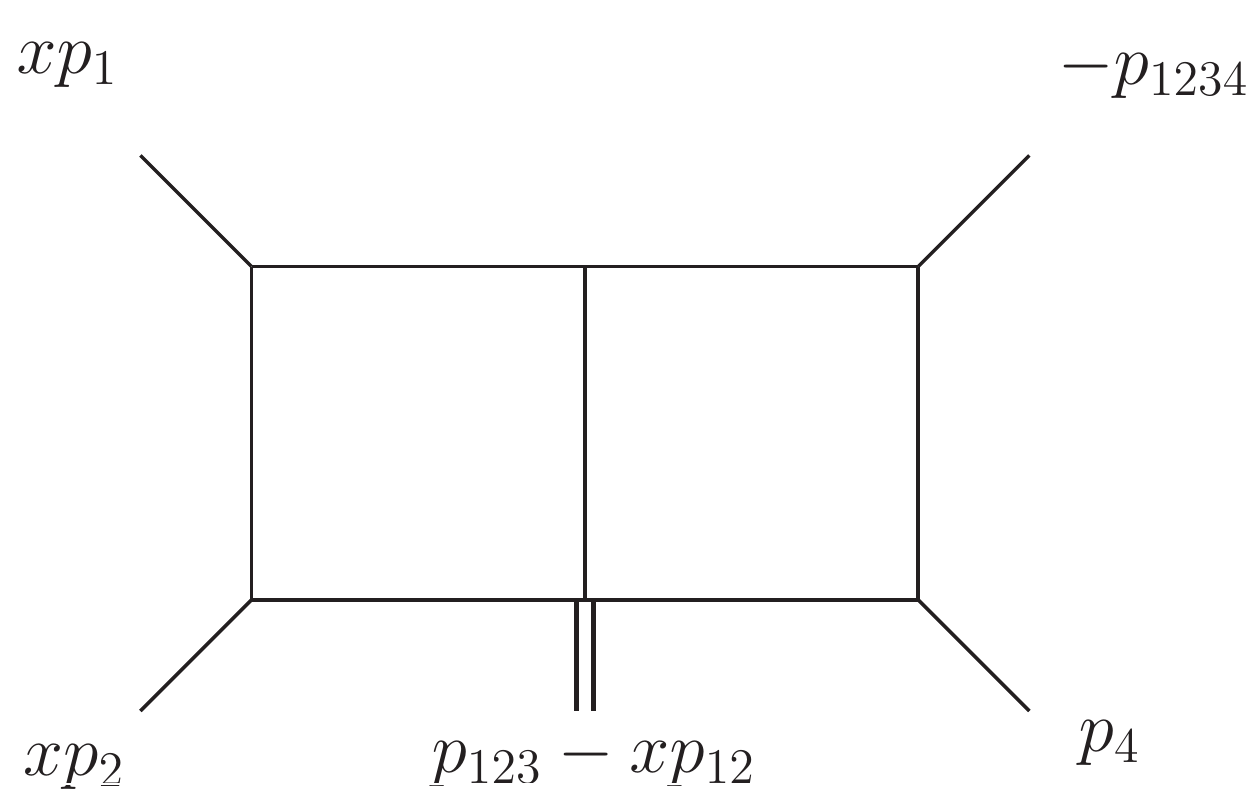} \hspace{0.4 cm}
\includegraphics[width=0.22 \linewidth]{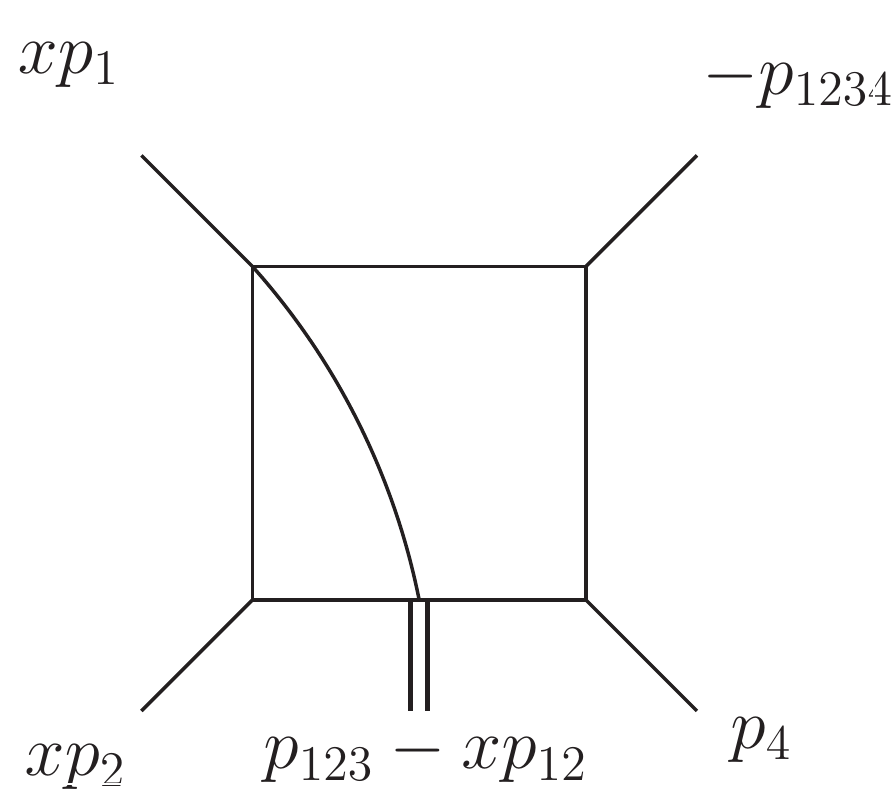} \\
\includegraphics[width=0.29 \linewidth]{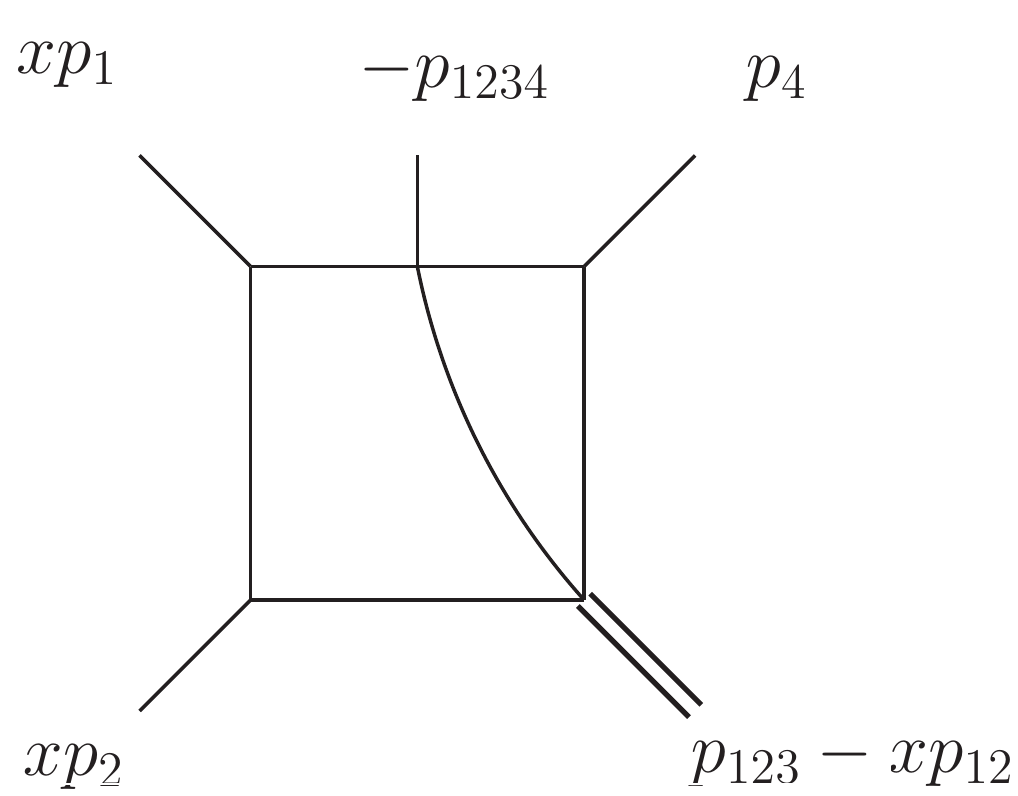} \hspace{0.3 cm}
\includegraphics[width=0.29 \linewidth]{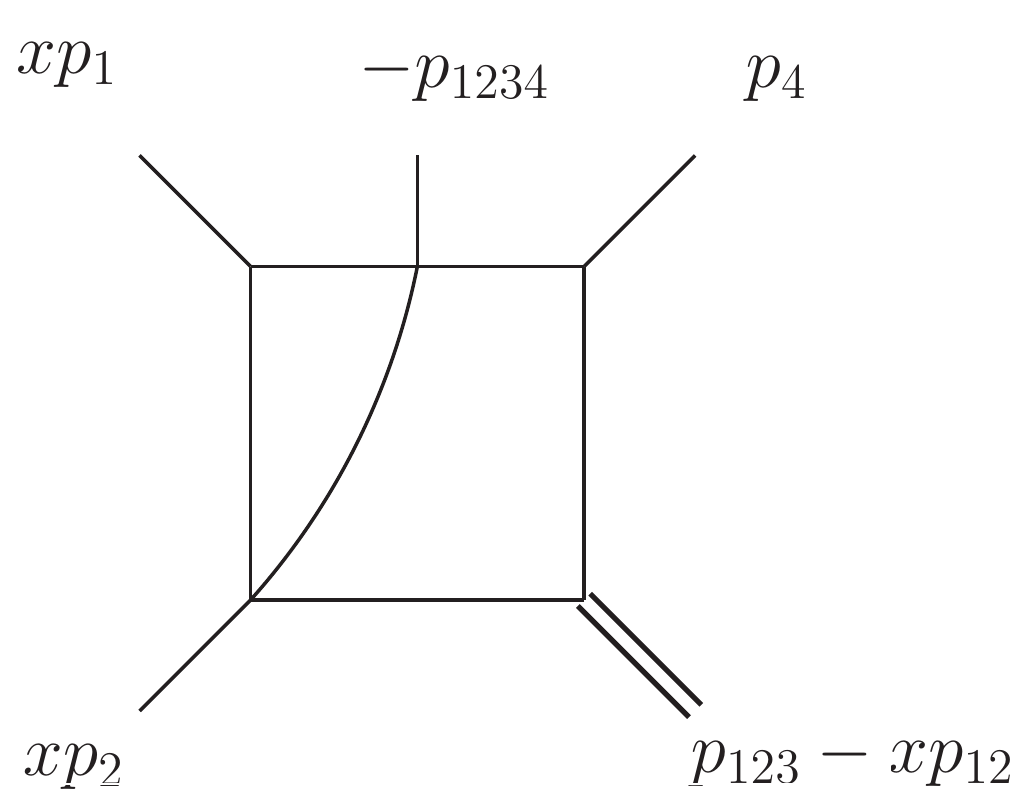} \hspace{0.3 cm}
\includegraphics[width=0.27 \linewidth]{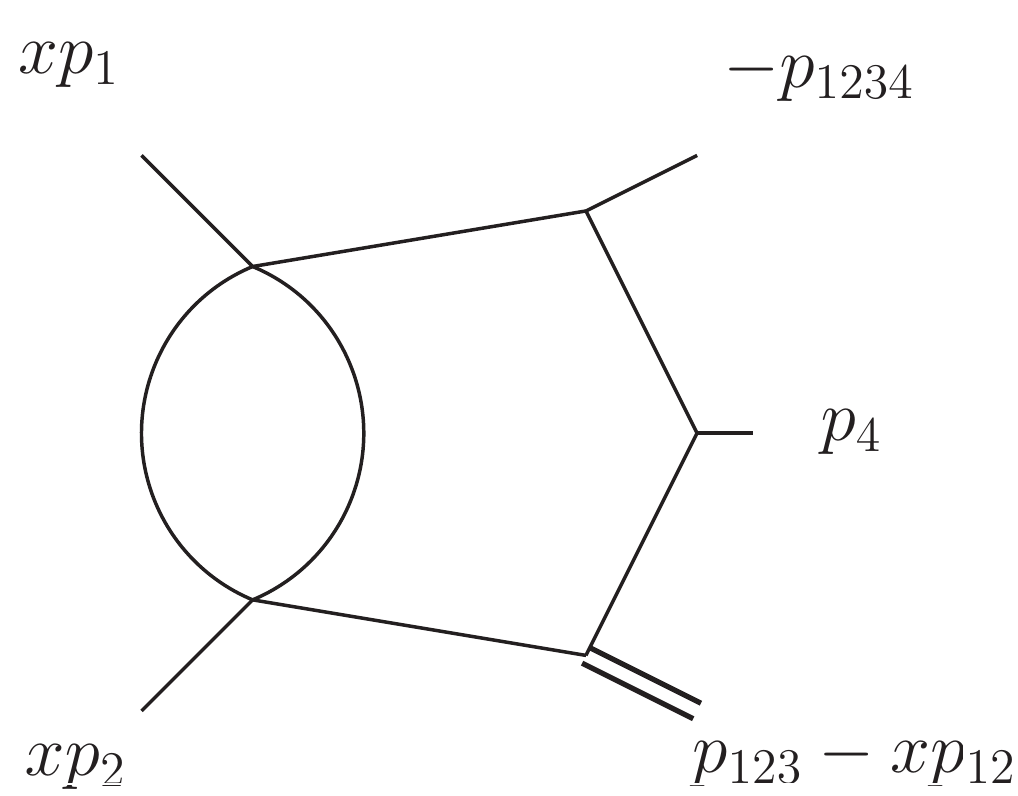}
  \caption{The five-point Feynman diagrams, besides the pentabox itself in Figure \ref{fig:param-P}, that are contained in the family $P_1$. All external momenta are incoming.}
  \label{fig:5point}
\end{figure}

The family $P_1$ contains in total 74 MI. Up to five denominators integrals with doubled propagators have been used as MI whereas starting from six denominators integrals with irreducible numerators  are chosen.  Many of the 74 MI already appear in the families of the  double box integrals discussed in~\cite{Gehrmann:2000zt,Gehrmann:2001ck,Gehrmann:2013cxs,Henn:2014lfa,Gehrmann:2014bfa,Caola:2014lpa,Papadopoulos:2014hla,Gehrmann:2015ora}. However, there are seventeen new five-point Feynman diagrams that are not contained in the double box integral families. Three of them are pentaboxes, including the scalar and two MI with irreducible numerators. 
There are six seven-denominator, and eight six-denominator ones, the scalar members of which are shown in Figure \ref{fig:5point}. 


\begin{figure}[t!]
\centering
\includegraphics[width=0.35 \linewidth]{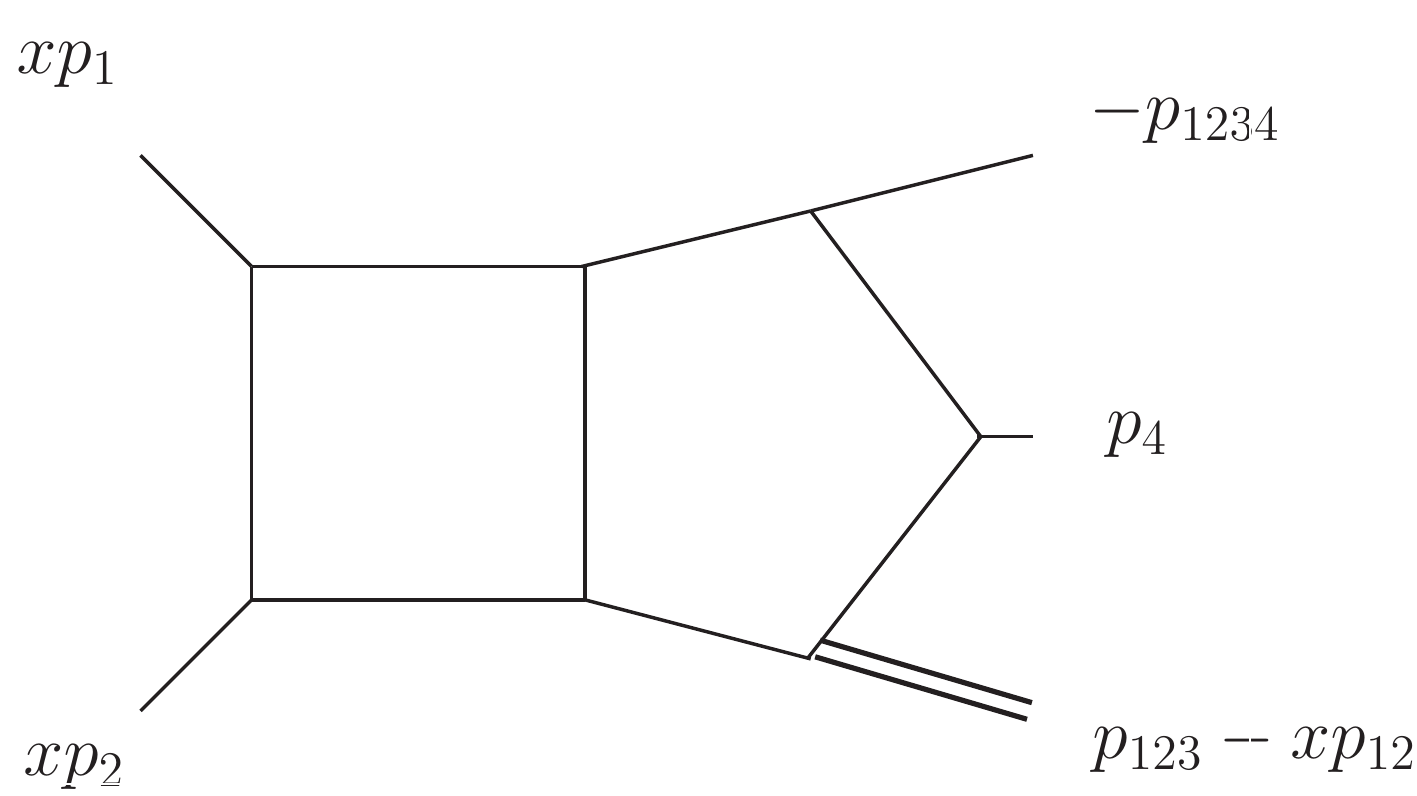}
  \caption{The parametrization of external momenta in terms of $x$ for the planar pentabox of the family $P_1$. All external momenta are incoming.}
  \label{fig:xparam-P1}
\end{figure}

For the family of integrals $P_1$ the external momenta are parametrized in $x$ as shown in Figure \ref{fig:xparam-P1}. The parametrization is chosen such that the double box MI with two massive external legs that is contained in the family $P_1$ has exactly the same parametrization as that one chosen in~\cite{Papadopoulos:2014hla}, i.e. two massless external momenta $xp_1$ and $xp_2$ and two massive external momenta $p_{123}-xp_{12}$ and $-p_{123}$. The MI in the family $P_1$ are therefore a function of a parameter $x$ and the following five invariants:
\begin{gather}
s_{12}:=p_{12}^2, \hspace{0.5 cm} s_{23}:=p_{23}^2, \hspace{0.5 cm} s_{34}:=p_{34}^2, \hspace{0.5 cm} s_{45}:=p_{45}^2=p_{123}^2, \hspace{0.5 cm} s_{51}:=p_{15}^2=p_{234}^2, \hspace{0.5 cm} p_i^2=0, 
\label{eq:qxpar}
\end{gather}
where the notation $p_{i\cdots j}=p_i+\cdots +p_j$ is used and $p_5:=-p_{1234}$. As the parameter $x\rightarrow 1$, the external momentum $q_3$ becomes massless, such that our parametrization (\ref{eq:qxpar}) also captures the on-shell case $M_3^2\rightarrow 0$. 

The MI depend in total on 6 variables, namely the Lorentz products $q_i.q_j$ with $i<j<5$ and the (squared) particle mass $M_3^2=q_3^2$. The $x$-parameterization of the external momenta as in Figure \ref{fig:xparam-P1} results in these variables being related to the parameter $x$ and the five independent scalar products of our choice that are defined in (\ref{eq:qxpar}). The  momenta $q_1,q_2,q_4$ and $q_5$ of the massless external particles can correspond to either of the four massless external legs in Figure \ref{fig:xparam-P1}, while the massive particle $V$ has an external momentum $q_3=p_{123}-xp_{12}$ with a mass:
\begin{gather}
M_3^2=(1 - x) (s_{45} - s_{12} x).\label{eq:Pinv}
\end{gather}

After fixing the $x$-parameterization as in Figure \ref{fig:xparam-P1}, the class of loop integrals describing the planar family
$P_1$ is now explicitly expressed in $x$ as:
\begin{gather}
G^{P_1}_{a_1\cdots a_{11}}(x,s,\epsilon):=e^{2\gamma_E \epsilon} \int \frac{d^dk_1}{i\pi^{d/2}}\frac{d^dk_2}{i\pi^{d/2}}
\frac{1}{k_1^{2a_1} (k_1 + x p_1)^{2a_2} (k_1 + x p_{12})^{2a_3} (k_1 + p_{123})^{2a_4}} \nonumber\\
\times \frac{1}{(k_1 + p_{1234})^{2a_5} k_2^{2a_6} (k_2 - x p_1)^{2a_7} (k_2 - x p_{12})^{2a_8} 
(k_2 - p_{123})^{2a_9} (k_2 - p_{1234})^{2a_{10}} (k_1 + k_2)^{2a_{11}}}, \label{eq:P1x}
\end{gather}
where $\gamma_E$ is the usual Euler-Mascheroni constant.

Using the notation given in Eq. (\ref{eq:P1x}), the indices $a_1\cdots a_{11}$ for the list of MI in the planar family $P_1$ is as follows\footnote{The letter m is used here to indicate the index $-1$.}:
{\footnotesize
\begin{eqnarray}
P_1: \hspace{-0.1cm} &\{10000000101, 01000000101, 00100000101, 10000001001,01000000011, 00100000011, 10100001100, \nonumber\\
&10100001010,10100101000, 01000101001, 10100100100, 10100000102,10100000101, 10100000011, \nonumber\\
&10000001102, 10000001101, 10000001011, 01000100101, 01000001101, 01000001011,00100100102, \nonumber\\
& 00100100101, 11100000101, 11100000011,11000001102, 11000001101, 11000001012, 11000001011, \nonumber\\
&11000000111, 10100000112, 10000001111, 01100100102, 01100100101, 01100100011, 01100000111, \nonumber\\
&01000101102,01000101101, 01000101011, 01000100111, 01000001111, 00100100111, 10100101100, \nonumber\\
&10100100101, 10100001101,10100001011, 10100000111, 111\text{m}0000111, 110000\text{m}1111, 11000001111, \nonumber\\
&10100101110, 10100100111, 10100001111,011001\text{m}0111, 01100100111, 010\text{m}0101111, 01000101111, \nonumber\\
&11100100101, 11100001101, 11100001011, 11100000111,111\text{m}0101101, 111001\text{m}1101, 11100101101, \nonumber\\
&1110\text{m}101011, 11100101011, 111\text{m}0100111, 11100100111, 111000\text{m}1111,111\text{m}0001111, 11100001111, \nonumber\\
&111001\text{m}0111, 11100101111, 111001\text{m}1111, 111\text{m}0101111\}, 
\label{eq:MIP1}
\end{eqnarray}
}
In the next section we discuss the DE method that we use to calculate the above 74 MI in $P_1$.

\section{Differential equations and their solution}
\label{sec:diff}


The resulting differential equation in matrix form can be written as 
\begin{equation}
{\partial _x}{\bf{G}} = {\bf{M}}\left( {\left\{ {{s_{ij}}} \right\},\varepsilon ,x} \right){\bf{G}}
\label{eq:ini}
\end{equation}
where ${\bf{G}}$ stands for the array of the 74 MI given in Eq.(\ref{eq:MIP1}).  
The diagonal part of the matrix at $\varepsilon=0$, namely ${\left( {{M_D}} \right)_{IJ}} = {\delta _{IJ}}\,{M_{II}}\left( {\varepsilon  = 0} \right), I,J=1\ldots 74$, defines as usual the integrating factors,  
and the equation takes the form
${\partial _x}{\bf{G}} = {\bf{MG}}$ with ${\bf{G}} \to  {{\bf{S}}^{ - 1}}{\bf{G}}$,  ${\bf{S}} = \exp \left( {\int {dx\,{{\bf{M}}_D}} } \right)$ and ${\bf{M}} \to {{\bf{S}}^{{\bf{ - 1}}}}\left( {{\bf{M - }}\,{{\bf{M}}_D}} \right){\bf{S}}$.

We found that, after absorbing the integrating factors, the resulting matrix $\bf{M}$ can be written as 
\begin{equation}
{M_{IJ}} = {N_{IJ}}\left( \varepsilon  \right)\left( {\sum\limits_{i = 1}^{20} {\sum\limits_{j = 1}^2 {\sum\limits_{k = 0}^1 {\frac{{{C_{IJ;ijk}}{\varepsilon ^k}}}{{{{\left( {x - {l_i}} \right)}^j}}}} } }  + \sum\limits_{j = 0}^1 {\sum\limits_{k = 0}^1 {{{\tilde C}_{IJ;jk}}{\varepsilon ^k}{x^j}} } } \right).
\label{eq:mprime}
\end{equation}
where the coefficients $C_{IJ;ijk}$ and ${\tilde C}_{IJ;jk}$ are rational functions of the kinematics $\{ s_{ij} \}$ and $ {N_{IJ}}\left( \varepsilon  \right)$ are rational function of $\varepsilon$ .
The twenty letters $l_i$, are given explicitly by
\begin{eqnarray}
%
& 0,\,\,\,1,\,\,\,\,\,\frac{{{s_{45}}}}{{{s_{45}} - {s_{23}}}},\,\,\,\,\,\frac{{{s_{45}}}}{{{s_{12}}}},\,\,\,\,\,1 - \frac{{{s_{34}}}}{{{s_{12}}}},\,\,\,\,\,1 + \frac{{{s_{23}}}}{{{s_{12}}}}, \nonumber\\
& 1 - \frac{{{s_{34}} - {s_{51}}}}{{{s_{12}}}},\,\,\,\,\,\frac{{{s_{45}} - {s_{23}}}}{{{s_{12}}}},\,\,\,\, - \frac{{{s_{51}}}}{{{s_{12}}}},\,\,\,\,\,\frac{{{s_{45}}}}{{ - {s_{23}} + {s_{45}} + {s_{51}}}},\,\,\,\,\,\frac{{{s_{45}}}}{{{s_{34}} + {s_{45}}}},
\nonumber
\\
& \frac{{{s_{12}}{s_{23}} - 2{s_{12}}{s_{45}} - {s_{12}}{s_{51}} - {s_{23}}{s_{34}} + {s_{34}}{s_{45}} - {s_{45}}{s_{51}} \pm \sqrt {{\Delta _1}} }}{{2{s_{12}}({s_{23}} - {s_{45}} - {s_{51}})}},\,\,\,\,\,\frac{{{s_{12}}{s_{23}} - {s_{12}}{s_{45}} - {s_{12}}{s_{51}} - {s_{23}}{s_{34}} + {s_{34}}{s_{45}} - {s_{45}}{s_{51}} \pm \sqrt {{\Delta _2}} }}{{2{s_{12}}({s_{23}} - {s_{45}} - {s_{51}})}},
\nonumber
\\
& \frac{{{s_{12}}{s_{23}} - {s_{12}}{s_{51}} - {s_{23}}{s_{34}} + {s_{34}}{s_{45}} - {s_{45}}{s_{51}} \pm \sqrt {{\Delta _1}} }}{{2{s_{12}}({s_{23}} + {s_{34}} - {s_{51}})}},\,\,\,\,\,\,\,\,\,\frac{{{s_{12}}{s_{45}} \pm \sqrt {{\Delta _3}} }}{{{s_{12}}{s_{34}} + {s_{12}}{s_{45}}}},\,\,\,\,\,\,\,\,\frac{{{s_{45}}}}{{{s_{12}} + {s_{23}}}},
\label{eq:lett}
\end{eqnarray}
where
\[\begin{array}{l}
{\Delta _1} = {({s_{12}}({s_{51}} - {s_{23}}) + {s_{23}}{s_{34}} + {s_{45}}({s_{51}} - {s_{34}}))^2} + 4{s_{12}}{s_{45}}{s_{51}}({s_{23}} + {s_{34}} - {s_{51}})\\
{\Delta _2} = {({s_{{\rm{12}}}}( - {s_{{\rm{23}}}} + {s_{{\rm{45}}}} + {s_{{\rm{51}}}}) + {s_{{\rm{23}}}}{s_{{\rm{34}}}} + {s_{{\rm{45}}}}({s_{{\rm{51}}}} - {s_{{\rm{34}}}}))^2} - 4{s_{{\rm{12}}}}{s_{{\rm{45}}}}{s_{{\rm{51}}}}( - {s_{{\rm{23}}}} + {s_{{\rm{45}}}} + {s_{{\rm{51}}}})\\
{\Delta _3} =  - ({s_{12}}{s_{34}}{s_{45}}({s_{12}} - {s_{34}} - {s_{45}}))
\end{array}\]
with $\Delta_1$ being the usual Gram determinant. The normalization factors $N_{IJ}\left( \varepsilon  \right)$ can be cast in the factorized form 
$ {N_{IJ}}\left( \varepsilon  \right) = {n_J} \left( \varepsilon  \right) /{n_I}\left( \varepsilon  \right) $ and can be absorbed by redefining ${{G}_I} \to {n_I}\left( \varepsilon  \right){{G}_I}$.

Although the DE can be solved starting from (\ref{eq:mprime})\footnote{Notice that terms corresponding to $k=0$ in (\ref{eq:mprime}) appear only  in non-diagonal triangular form.} and the result can be expressed as a sum of GPs with argument $x$ and weights given by the letters in Eq. (\ref{eq:lett}), it is more elegant and easier to solve, if the system of differential equations
is brought into a Fuchsian form~\cite{Henn:2014qga}, where only single poles in the variable $x$ appear. In fact the series of successive transformations
\begin{equation}
\label{eq:trans}
{\bf{G}} \to \left({\bf{I}}-{\bf{K}}_i\right){\bf{G}}, \quad {\bf{M}} \to \left( {{\bf{M}} - {\partial _x}{{\bf{K}}_i} - {{\bf{K}}_i}{\bf{M}}} \right){\left( {{\bf{I}} - {{\bf{K}}_i}} \right)^{ - 1}}\,\,\,\,i = 1,2,3
\end{equation}
with 
\[{\left( {{{\bf{K}}_1}} \right)_{IJ}} = \left\{ {\begin{array}{*{20}{c}}
{\int {dx} {{\left( {{\bf{M}}\left( {\varepsilon  = 0} \right)} \right)}_{IJ}}}&{\,I,J \ne 69,74}\\
0&{I,J = 69,74}
\end{array}} \right.\]
\[{\left( {{{\bf{K}}_2}} \right)_{IJ}} = \left\{ {\begin{array}{*{20}{c}}
{\int {dx} {{\left( {{\bf{M}}\left( {\varepsilon  = 0} \right)} \right)}_{IJ}}}&{\,I,J \ne 74}\\
0&{I,J = 74}
\end{array}} \right.\]
and 
\[{\left( {{{\bf{K}}_3}} \right)_{IJ}} = \int {dx} {\left( {{\bf{M}}\left( {\varepsilon  = 0} \right)} \right)_{IJ}}\]
with the enumeration of the MI as given by Eq. (\ref{eq:MIP1}), brings the system into the form
\begin{equation}
{\partial _x}{\bf{G}} = \left( {\varepsilon \sum\limits_{a = 1}^{19} {\frac{{{{\bf{M}}_a}}}{{\left( {x - {l_a}} \right)}}} } \right){\bf{G}}
\label{eq:fuchsian}
\end{equation}
where the residue matrices ${{{\bf{M}}_a}}$ are independent of $x$ and $\varepsilon$.

The rationale of the above transformations is of course to bring the system in its Fuchsian form. Starting from the matrix in (\ref{eq:mprime}) and after the absorption of the
$\varepsilon$-dependent terms  $N_{IJ}\left( \varepsilon  \right)$, the matrix at order $\varepsilon^0$, with the exception of the rows $69$ and $74$,
only contains terms proportional to $(x-l_i)^{-2}$ and $x^0$, where $l_i$ are letters from (\ref{eq:lett}). The transformation based on  ${\bf K}_1$ matrix, whose elements are proportional to
 $(x-l_i)^{-1}$ and $x^1$, obviously aims at removing these terms.
As a matter of fact the new matrix, after the first transformation is applied, has, at order $\varepsilon^0$, non-zero elements only in the rows $69$ and $74$, again proportional to $(x-l_i)^{-2}$ and $x^0$, 
with the exception of the row $74$. Finally, after the
second transformation, the whole matrix at $\varepsilon^0$-order has only non-zero elements in the row $74$ proportional to $(x-l_i)^{-2}$ and $x^0$, so by the application of ${\bf K}_3$ transformation, 
the system is brought into the form (\ref{eq:fuchsian}). 
 It should be noticed that the series of the above transformations {\it do not} correspond to the one described by the Moser 
algorithm~\cite{Lee:2014ioa,BarkatouPfluegel2009,BarkatouPfluegel2007,Moser1960}. 
Moreover, the final form does not only correspond to a Fuchsian form, but also factors out completely the  $\varepsilon$ dependence.

Equation (\ref{eq:fuchsian}) can be straightforwardly solved and the result is given as 
\begin{eqnarray}
\label{eq:solx}
{\bf{G}} &=& {\varepsilon ^{ - 2}}{\bf{b}}_0^{\left( { - 2} \right)} + {\varepsilon ^{ - 1}}\left( {\sum {{{\cal G}_a}{{\bf{M}}_a}{\bf{b}}_0^{\left( { - 2} \right)}}  + {\bf{b}}_0^{\left( { - 1} \right)}} \right) 
\nonumber \\
&+& {\varepsilon ^{0}} \left( {\sum {{{\cal G}_{ab}}{{\bf{M}}_a}{{\bf{M}}_b}} {\bf{b}}_0^{\left( { - 2} \right)} + \sum {{{\cal G}_a}{{\bf{M}}_a}{\bf{b}}_0^{\left( { - 1} \right)}}  + {\bf{b}}_0^{\left( 0 \right)}} \right)
\nonumber \\
&+& \varepsilon \left( {\sum {{{\cal G}_{abc}}{{\bf{M}}_a}{{\bf{M}}_b}{{\bf{M}}_c}} {\bf{b}}_0^{\left( { - 2} \right)} + \sum {{{\cal G}_{ab}}{{\bf{M}}_a}{{\bf{M}}_b}{\bf{b}}_0^{\left( { - 1} \right)}}  + \sum {{{\cal G}_a}{{\bf{M}}_a}{\bf{b}}_0^{\left( 0 \right)} + } {\bf{b}}_0^{\left( 1 \right)}} \right)
\nonumber \\
&+& {\varepsilon ^2}\left( {\sum {{{\cal G}_{abcd}}{{\bf{M}}_a}{{\bf{M}}_b}{{\bf{M}}_c}{{\bf{M}}_d}} {\bf{b}}_0^{\left( { - 2} \right)} + \sum {{{\cal G}_{abc}}{{\bf{M}}_a}{{\bf{M}}_b}{{\bf{M}}_c}{\bf{b}}_0^{\left( { - 1} \right)}} } \right.
\nonumber \\
&+& \left. {\sum {{{\cal G}_{ab}}{{\bf{M}}_a}{{\bf{M}}_b}{\bf{b}}_0^{\left( 0 \right)} + \sum {{{\cal G}_a}{{\bf{M}}_a}{\bf{b}}_0^{\left( 1 \right)}}  + } {\bf{b}}_0^{\left( 2 \right)}} \right)
\end{eqnarray}
with the arrays ${\bf{b}}_0^{\left( {k} \right)}$, $k=-2,...,2$ representing the $x$-independent boundary terms in the limit $x\to 0$ at order 
$\varepsilon^k$, 
\[{\bf{G}}\mathop  \sim \limits_{x \to 0} \sum\limits_{k =  - 2}^2 {{\varepsilon ^k}\,\sum\limits_{n = 0}^{k + 2} {{\bf{b}}_n^{\left( k \right)}{{\log }^n}\left( x \right)} }  + {\mathrm{subleading\;  terms}}.\]

The expression (\ref{eq:solx}) is in terms of Goncharov polylogarithms,
${{\cal G}_{a,b, \ldots }} = {\cal G}\left( {{l_a},{l_b}, \ldots ;x} \right)$ with $a,b,c,d=1,\ldots,19$.
In the 
literature\footnote{See for instance references~\cite{Henn:2013pwa,Henn:2014qga}.}, the so-called canonical d-log form of a system of DE, is defined as 
$d{\bf{G}}\left( {\vec x,\varepsilon } \right) = \left( {\varepsilon \sum\limits_k {{{\bf{A}}_k}d\log {\alpha _k}\left( {\vec x} \right)} } \right){\bf{G}}\left( {\vec x,\varepsilon } \right)$, 
with ${\alpha _k}\left( {\vec x} \right)$, the so-called alphabet,  being rational functions of the independent variables $\{\vec x\}$.
Although this form looks very similar to our equation (\ref{eq:fuchsian}), with the identification ${\alpha _k}\left( x \right) := x - {l_k}$, a comment 
is in order: usually in the canonical d-log form, DE are constructed with respect to all kinematical variables, and therefore the corresponding matrices $\mathbf{A}_k$ should not depend on the kinematics.
Contrary in our case, since the DE are derived with respect to {\it only one} variable, namely $x$, the matrices $\mathbf{M}_a$ still depend on the (rest of the) kinematics $\{s_{ij}\}$, defined in (\ref{eq:qxpar}). 
As a consequence, in the multi-variable DE, the results are expressed
in terms of GPs with coefficients that are rational numbers -- commonly referred as universally transcendental (UT) -- whereas in our case coefficients of GPs may still depend on $\{s_{ij}\}$.
In fact most of the elements of ${\mathbf G}$ 
in our solution do 
appear in a form that corresponds to the commonly defined UT form, with coefficients of GPs that are rational numbers independent of $\{s_{ij}\}$, 
though others exhibit coefficients of GPs depending on $\{s_{ij}\}$. In fact, working with DE in one independent variable, 
one can easily accommodate the dependence on the kinematics, as shown explicitly in Eq.(\ref{eq:solx}),
although it is an open question if the DE can be cast into a form so that all residue matrices $\mathbf{M}_a$ are independent of the kinematics.
In this context it is also interesting to mention that all the residue matrices $\mathbf{M}_a$, with a non-trivial dependence on the kinematics, do have eigenvalues that are negative integer numbers independent of the kinematics,  
related to the behaviour ${\left( {x - {l_a}} \right)^{ - {n_a}\varepsilon }}$($n_a$ positive integers) of the integrals at the corresponding limits $x\to l_a$.

The limit $x=1$ represents the solution for all planar pentabox on-shell Feynman integrals. The limit can easily be obtained by properly resumming the ${\log ^k}\left( {1 - x} \right)$ terms. Interestingly enough we found a very simple formula 
for this limit given by 
\begin{equation}
{{\bf{G}}_{x = 1}} = \left( {{\bf{I}} + \frac{3}{2}{{\bf{M}}_2} + \frac{1}{2}{\bf{M}}_2^2} \right){{\bf{G}}_{trunc}}
\label{eq:limit}
\end{equation}
with ${\bf{M}}_2$ the residue matrix at $x=1$ and ${{\bf{G}}_{trunc}}$ derived from Eq.(\ref{eq:solx}), by properly removing all divergencies proportional to ${\log ^k}\left( {1 - x} \right)$ and setting $x=1$. Details can be found in Appendix \ref{x=1}.

For the majority of the MI in the original basis (\ref{eq:MIP1}), their boundary behaviour is captured by the DE itself, as it was also the case for the doublebox families~\cite{Papadopoulos:2014hla}.
This is achieved in a  {\it bottom-up} approach. 
For several MI, such as the one-scale MI, including the genuine two-loop MI with three denominators,  
as well as for those MI that are expressed as product of one-loop integrals, their dependence on $x$ is known in a closed form
in terms proportional to $x^{i+j\epsilon}$. 
The rest of the MI, with a number $m$ of denominators ($m>3$), satisfy inhomogeneous DE, where the inhomogeneous part is
completely fixed by MI with a number $m^\prime$ of denominators, with $m^\prime < m$. 
Then, at each step of the iteration, we first calculate the  part of the corresponding MI with $m$ denominators, which we call the 
resummed part~\footnote{The expression of the MI in the limit $x\to 0$ before the limit $\varepsilon \to 0$ is considered.}, $G^{(m)}_{res}$,
by introducing 
the following parametrization 
\begin{equation}
G^{(m)}_{res}=\sum_j c_jx^{i_0+j\epsilon}+d_jx^{i_0+1+j\epsilon}
\label{eq:exp}
\end{equation}
where $i_0$ and $j$ are dictated by the DE itself in the limit $x\to 0$.
By putting the above expression (\ref{eq:exp}) for the integrals in the DE and equating the coefficients of the terms $x^{i+j\epsilon}$ with the same exponents on both sides of the DE in the limit $x\to 0$, 
linear equations are obtained for the coefficients $c_i$ and $d_i$.
We note here that for the large majority of integrals we did not need to solve for the coefficients $d_i$ that correspond to $x$-suppressed terms. Their calculation were only required for those integrals whose DE had singularities of the form 
$x^{-2+j\epsilon}$ at the boundary $x=0$ (such singularities were also encountered for the DE of the one-loop pentagon discussed in~\cite{Papadopoulos:2014lla}). Once the resummed terms in equation (\ref{eq:exp}) were calculated, the resulting 
DE for $G^{(m)}_{fin}:=G^{(m)}-G^{(m)}_{res}$ have no singularities at $x=0$ 
and therefore can be safely expanded in $\varepsilon$ and the result directly expressed in terms of Goncharov polylogarithms of the form ${\cal G}\left( {{l_a},{l_b}, \ldots ;x} \right)$.

For the pentabox family $P_1$ specifically, the majority of the coefficients (\ref{eq:exp}) are fixed by the above-mentioned linear equations, while some others are not. We found in practice that for most of the integrals, the coefficients which are not fixed by the linear equations are zero and we confirmed this by the method of expansion by regions~\cite{Beneke:1997zp,Smirnov:2002pj}. However, for some integrals we found that the method of expansion by regions predicts that some coefficients that are {\it not} determined by the linear equations, are in fact non-zero and require an explicit calculation. As described in the Appendix \ref{expbyreg}, for those integrals we used other methods to calculate the unknown and nonvanishing coefficients. Once all boundary conditions were found for the integrals in the original basis (\ref{eq:MIP1}), the boundary conditions for the canonical basis followed directly from the relation between the two bases that is described above.

It should be noticed that in the language of expansion by regions~\cite{Beneke:1997zp,Smirnov:2002pj}, MI in the limit $x\to 0$, 
can be written, in general, as an expansion in terms proportional to $x^{i+j\epsilon}\log^k(x)$, with $k \ge 0$, which goes beyond our parametrization in (\ref{eq:exp}). Nevertheless, we have explicitly checked,
either by using the method of  expansion by regions~\cite{Jantzen:2012mw} or by comparing with previously known results, that
in our case, for the pentabox family $P_1$, no terms with $k\ge 1$ were present, at the leading order in the limit $x\to 0$.

The complete expressions for all MI are available in the ancillary files~\cite{results}. The solution for all 74 MI contains $O(3,000)$ GPs which is approximately six times more than the corresponding double-box with two off-shell legs planar MI.
 We have performed several numerical checks of all our calculations. 
The numerical results, also included in the ancillary files~\cite{results}, have been obtained with the {\bf GiNaC} library~\cite{Vollinga:2004sn} and compared with those provided by the numerical code {\bf SecDec}~\cite{Binoth:2000ps,Heinrich:2008si,Borowka:2012yc,Borowka:2013cma,Borowka:2015mxa} in the Euclidean region for all MI and in the physical region whenever possible
(due to CPU time limitations in using {\bf SecDec}) and perfect agreement was found. For the physical region we are using the analytic continuation as described in~\cite{Papadopoulos:2014hla}.
At the present stage we are not setting a fully-fledged numerical implementation, which will be done when all families will be computed. Our experience with double-box computations show that using for instance {\tt HyperInt}~\cite{Panzer:2014caa}
 to bring all GPs in their range of convergence, {\it before} evaluating them numerically by {\bf GiNaC}, 
increases efficiency by two orders of magnitude.  Moreover expressing GPs in terms of classical polylogarithms and $Li_{2,2}$, could also reduce substantially the CPU time~\cite{Gehrmann:2015ora}. 
Based on the above we estimate that a target of  $O\left( {{{10}^2-{10}^3}} \right)$ milliseconds can be achieved.

\section{Conclusions}

In this paper we calculated, for the first time, one of the topologies of planar Master Integrals related to massless five-point amplitudes with one off-shell leg as well as the full set of massless planar Master Integrals for on-shell kinematics. 
We have demonstrated that based on the Simplified Differential Equations approach~\cite{Papadopoulos:2014lla} these MI can be expressed in terms of Goncharov polylogarithms. The complexity of the resulting expressions is certainly promising that 
the project of computing all MI relevant to massless QCD, namely all eight-denominator MI with arbitrary configuration of the external momenta, is feasible. Having such a complete library of two-loop MI, the analog of $A_0, B_0, C_0, D_0$ scalar integrals at one loop,  the reduction of an arbitrary two-loop amplitude {\`a la OPP} can pave the road  for a NNLO automation in the near future.

As experience shows, there are several issues that will need to attract our attention in order to accomplish our goal. First of all in order to systematize the whole procedure of reducing an arbitrary Feynman Integral in terms of MI in an efficient way, a deepening of our current understanding of 
IBP identities~\cite{Baikov:1996iu,Smirnov:2003kc}
is necessary~\cite{Ita:2015tya}. Secondly, further standardising the procedure to obtain a canonical form of DE~\cite{Henn:2013pwa}, which drastically simplifies the expression of MI in terms of GPs, is certainly a very desirable feature. Thirdly, the inclusion of MI with massive
internal propagators, at a first stage with one mass scale corresponding to the heavy top quark, will provide the complete basis for NNLO QCD automated computations. Moreover, the calculation of boundary terms for the DE can benefit from further 
developments and exploitations of expansion-by-regions techniques, in conjunction with Mellin-Barnes representation of the resulting integrals. Finally, on the numerical side, a more efficient computation of polylogarithms is also necessary.

\subsection*{Acknowledgements}

This research was supported by the Research Funding Program ARISTEIA, HOCTools (co-financed by the European Union (European Social Fund ESF) and Greek national funds through the Operational Program "Education and Lifelong Learning" of the National Strategic Reference Framework (NSRF)). 
We gratefully acknowledge fruitful discussions with J.~Henn, A.~V.~Smirnov, V.~A.~Smirnov,  M.~Czakon, S.~Weinzierl, C.~Duhr and E.~Panzer during various stages of this project. We would like also to thank H.~Frellesvig for providing us with useful symbolic code to manipulate
GPs, and A.~Kardos and G.~Somogyi for cross-checking some of the integrals appearing in the expansion-by-region method to compute boundary terms. We would like also to thank S.~Borowka, E.~Panzer and A.~V.~Smirnov for helping with
computer programs {\tt SecDec}, {\tt HyperInt} and {\tt FIRE} respectively. 

\appendix

\section{The $x=1$ limit}
\label{x=1}

The result given in (\ref{eq:solx}) can be re-written as an expansion in $\log(1-x)$, in the following form 

\begin{equation}
{\bf{G}} = \sum\limits_{n \ge  - 2} {{\varepsilon ^n}} \sum\limits_{i = 0}^{n + 2} {\frac{1}{{i!}}{\bf{c}}_i^{\left( n \right)}{{\log }^i}\left( {1 - x} \right)} 
\label{eq:logx=1}
\end{equation}
where all coefficients ${\bf{c}}_i^{\left( n \right)}$ are finite in the limit $x=1$. This can be straightforwardly achieved, starting from (\ref{eq:solx}) by transporting all letters $l=1$ of GPs to the right according to their known shuffle properties~\cite{Vollinga:2004sn}.
It is easy to see that 
\[{\bf{c}}_i^{\left( n \right)} = {{\bf{M}}_2}{\bf{c}}_{i - 1}^{\left( {n - 1} \right)}\,\,\,\,\,\,\,i \ge 1\]
where the matrix ${{\bf{M}}_{2}}$ corresponds to the letter $l=1$ and by definition the regular part of ${\bf G}$ at $x=1$ is given by
\[{{\bf{G}}_{reg}} = \sum\limits_{n \ge  - 2} {{\varepsilon ^n}{\bf{c}}_0^{\left( n \right)}}. \]

Since the characteristic polynomial of  ${\bf M}_2$ is given by $x^{61} (1+x)^9 (2+x)^4$
it is natural to make the following ansatz for the resummation of $\log(1-x)$ terms
\begin{equation}
{\bf{G}} = {{\bf{G}}_{reg}} + \frac{{\left( {{{\left( {1 - x} \right)}^{ - 2\varepsilon }} - 1} \right)}}{{\left( { - 2\varepsilon } \right)}}{\bf{X}} + \frac{{\left( {{{\left( {1 - x} \right)}^{ - \varepsilon }} - 1} \right)}}{{\left( { - \varepsilon } \right)}}{\bf{Y}}
\label{eq:logx=1r}
\end{equation}
with 
\[{\bf{X}} = \sum\limits_{n \ge  - 1} {{\varepsilon ^n}{{\bf{X}}^{\left( n \right)}}} \,\,\,\,\,{\bf{Y}} = \sum\limits_{n \ge  - 1} {{\varepsilon ^n}{{\bf{Y}}^{\left( n \right)}}} .\]

By equating the powers of $\log(1-x)$ in the left-hand sides of (\ref{eq:logx=1}) and (\ref{eq:logx=1r}) we obtain towers of equations for the coefficients
${{\bf{X}}^{\left( n \right)}}$ and ${{\bf{Y}}^{\left( n \right)}}$. Consistency of these equations require the following identities to hold

\[{\left( { - 1} \right)^n}{{\bf{M}}_2}^n = {{\bf{M}}_2}^2\left( {{2^{n - 1}} - 1} \right) + {{\bf{M}}_2}\left( {{2^{n - 1}} - 2} \right),\;\;\;n\ge 1.\]

The above relation can  be  easily proved by mathematical induction. The first non-trivial relation corresponds to $n=3$, namely  $- {{\bf{M}}_2}^3 = 3{{\bf{M}}_2}^2 + 2{{\bf{M}}_2}$ which is a direct consequence of 
the minimal polynomial $x(x+1)(x+2)$ of the matrix ${\bf M}_2$. Equation (\ref{eq:limit}) can now be easily derived from (\ref{eq:logx=1r}) by taking the limit $x=1$ with ${\bf G}_{trunc}\equiv {\bf G}_{reg}(x=1)$.
 
\section{Methods of calculating the boundary conditions}
\label{expbyreg}

As was mentioned in section \ref{sec:diff}, for some integrals their DE do not completely fix their behaviour as $x\rightarrow x_0=0$. The indices of these integrals are:
{\footnotesize
\begin{gather}
\{(10100000101),(10100000102),(11000001012),(11000001011),(01000101011),(10100100111), \nonumber\\
(10100001111),(111\text{m}0100111),(111000\text{m}1111),(11100001111),(111001\text{m}0111),(10100000011) \nonumber\\
(10000001011),(01000001011),(11100000011),(01100100011),(10100100111),(11100001011), \nonumber\\
(11100101011),(111\text{m}0101011)\}.
\label{eq:boundnot}
\end{gather}
}
In other words, for the above integrals there is some behaviour at $x\rightarrow 0$ which corresponds to coefficients $c_i$ and $d_i$ in equation (\ref{eq:exp}) that are not all determined by the DE itself. For the above integrals we used various methods to compute the undetermined coefficients, which we explain further below.

\paragraph{Expansion by regions}

For the following integrals the method of expansion by regions was used to fix the non-zero coefficients in the boundary behaviour (\ref{eq:exp}) which are not determined by the DE:
{\footnotesize
\begin{gather}
\{(10100000101),(10100000102),(11000001012),(11000001011),(01000101011),(10100100111), \nonumber\\
(10100001111),(111\text{m}0100111),(111000\text{m}1111),(11100001111),(111001\text{m}0111)\}.
\label{eq:boundnotexp}
\end{gather}
}
With the method of expansion by regions~\cite{Beneke:1997zp,Smirnov:2002pj}, the coefficients in equation (\ref{eq:exp}) are expressed as Feynman integrals. For most of the integrals in (\ref{eq:boundnotexp}), these integrals corresponding to the undetermined (by the DE) coefficients are reducible to known single scale integrals. However for a few, the integrals for the coefficients are non-trivial and require further calculation. For example let's consider the following MI and its DE expanded around $x=0$:
\begin{gather}
G=G_{0100101011}=e^{2\gamma_E \epsilon}\int \frac{d^dk_1}{i\pi^{d/2}}\frac{d^dk_2}{i\pi^{d/2}}
\frac{1}{(k_1 + x p_1)^2 k_2^2 (k_2 - x p_{12})^2 (k_2 - p_{1234})^2 (k_1 + k_2)^2}, \nonumber\\
\partial_x(x^{1+4\epsilon}G)=B(s,\epsilon)x^{-1+2\epsilon}+\cdots, \label{eq:DEexp}
\end{gather}
where $B(s,\epsilon)$ is independent of the integral parameter $x$ and the dots do not contain any further poles at $x=0$. From expansion by regions it follows that at $x=0$ the integral behaves as $G\sim c_1x^{-1-4\epsilon}+c_2x^{-1-2\epsilon}$ which corresponds to a soft and collinear region. The coefficient $c_2$ is completely determined by the DE (\ref{eq:DEexp}) as explained in section \ref{sec:diff}, while $c_1$ is not. The coefficient $c_1$ is given by the method of regions as a Feynman integral:
\begin{equation}
c_1=e^{2\gamma_E \epsilon}\int \frac{d^dk_1}{i\pi^{d/2}}\frac{d^dk_2}{i\pi^{d/2}}
\frac{1}{(k_1 + p_1)^2 k_2^2 (k_2 - p_{12})^2 (- 2 k_2.p_{1234}) (k_1 + k_2)^2}.
\end{equation}
We can calculate the above integral by the method of DE. As we did for the $x$-parametrization of the $P_1$ family, we introduce a parameter $y$ in the denominators:
\begin{equation}
I(y):=e^{2\gamma_E \epsilon}\int \frac{d^dk_1}{i\pi^{d/2}}\frac{d^dk_2}{i\pi^{d/2}}
\frac{1}{(k_1 + y p_1)^2 k_2^2 (k_2 - p_{12})^2 (- 2 k_2.p_{1234}) (k_1 + k_2)^2}.
\end{equation}
By solving the DE $\partial_y I(y)$ and afterwards setting $y\rightarrow 1$ we could calculate $c_1=I(1)$. We note that the boundary condition for the integral $I(y)$ is completely determined by its DE. If this was not the case, we would use the method of regions again to express the missing coefficient in terms of a Feynman integral and repeat the above step, though this was never necessary.

\paragraph{Shifted boundary point}

For some other integrals we considered their limiting behaviour around another boundary point $x_0$ instead of at $x_0=0$:
{\footnotesize
\begin{equation}
\begin{array}{cc}
x_0=\infty: & \{(10100000011),(10000001011),(11100000011),(01100100011),(10100100111)\} \\
x_0=(s_{12} - s_{34} + s_{51})/s_{12}: & \{(01000001011)\}
\end{array}
\end{equation}
}
In practice, choosing the boundary at $x_0=\infty$ corresponds to performing an inverse transformation $x\rightarrow 1/x$ at the level of the DE and afterwards considering $x_0=0$. This case was already discussed in~\cite{Papadopoulos:2014hla} and therefore here we discuss the case of the MI $G:=G_{01000001011}$. Its DE is of the form:
\begin{gather}
\partial_x\left(\left(1 - \frac{s_{12} x}{s_{12} - s_{34}}\right)^{\epsilon} \left(1 - \frac{s_{12} x}{s_{12} - s_{34} + s_{51}}\right)^{\epsilon}x^{2\epsilon}G\right) \nonumber\\
=C(s,\epsilon)\left(1-\frac{s_{12} x}{s_{12}-s_{34}}\right)^{\epsilon -1} \left(1-\frac{s_{12} x}{s_{12}-s_{34}+s_{51}}\right)^{\epsilon -1},
\label{eq:bound2}
\end{gather}
where $C(s,\epsilon)$ is a factor independent of the integral parameter $x$. By using the method of expansion by regions, it can be checked that $(1 - \frac{s_{12} x}{s_{12} - s_{34}})^{\epsilon} (1 - \frac{s_{12} x}{s_{12} - s_{34} + s_{51}})^{\epsilon}x^{2\epsilon}G\rightarrow 0$ as $x\rightarrow (s_{12} - s_{34} + s_{51})/s_{12}$. The right hand side of equation (\ref{eq:bound2}) has a singularity at $x=(s_{12}-s_{34}+s_{51})/s_{12}$, which exactly captures the behaviour of the integral $G$ around $x=(s_{12}-s_{34}+s_{51})/s_{12}$. Therefore by the transformation $x\rightarrow \frac{s_{12} - s_{34} + s_{51}}{s_{12}}(1-x')$ and then afterwards integrating from $x'=0$ one can compute $G$ as a function of $x'$. Upon tranforming back to $x$, the limiting behaviour of $G$ at $x=0$ and therefore its corresponding boundary condition is found.

\paragraph{Extraction from known integrals}

The last three integrals in (\ref{eq:boundnot}) correspond to taking the $x\rightarrow 1$ limit of other known integrals (that lie in the same family $P_1$) and afterwards redefining the invariants as follows:
{\small
\begin{gather}
G_{11100001011}(x,s_{12},s_{34},s_{51})=G_{11100100101}(x'=1,s_{12}',s_{23}',s_{45}'), \nonumber\\
G_{11100101011}(x,s_{12},s_{34},s_{51})=G_{11100101101}(x'=1,s_{12}',s_{23}',s_{45}'), \nonumber\\
G_{111\text{m}0101011}(x,s_{12},s_{34},s_{51})=G_{111\text{m}0101101}(x'=1,s_{12}',s_{23}',s_{45}'), \nonumber\\
s_{12}'=x^2s_{12}, \hspace{0.5 cm} s_{23}'=x s_{51}, \hspace{0.5 cm} s_{45}'=-xs_{12} + xs_{34} + x^2s_{12}.
\label{eq:limits}
\end{gather}
}
The three integrals on the right hand side of equation (\ref{eq:limits}) are MI that were previously calculated in~\cite{Papadopoulos:2014hla}. From the exact result in $x$ for the three integrals on the left hand side in (\ref{eq:limits}) we could then compute their corresponding non-zero coefficients in (\ref{eq:exp}) that are not determined\footnote{For each of the three integrals, there was exactly one coefficient which was not fixed by the DE as explained in section \ref{sec:diff}.} by the DE.

\bibliographystyle{JHEP}
\bibliography{pentabox_paper}

\providecommand{\href}[2]{#2}\begingroup\raggedright\begin{thebibliography}{10}

\bibitem{Khachatryan:2015uqb}
{\bf CMS} Collaboration, V.~Khachatryan et~al., {\it {Measurement of the top
  quark pair production cross section in proton-proton collisions at
  $\sqrt{s}=13$ TeV}},  \href{http://xxx.lanl.gov/abs/1510.0530}{{\tt
  arXiv:1510.0530}}.

\bibitem{Atlas}
{\bf ATLAS} Collaboration, {\it {Measurement of the Production Cross Sections
  of a $Z$ Boson in Association with Jets in $pp$ collisions at $\sqrt{s} = 13$
  TeV with the ATLAS Detector}},  {\em ATLAS-CONF-2015-041} (2015).

\bibitem{Aad:2015nda}
{\bf ATLAS} Collaboration, G.~Aad et~al., {\it {Measurement of four-jet
  differential cross sections in $\sqrt{s}=8$ TeV proton--proton collisions
  using the ATLAS detector}},  \href{http://xxx.lanl.gov/abs/1509.0733}{{\tt
  arXiv:1509.0733}}.

\bibitem{Khachatryan:2015tzo}
{\bf CMS} Collaboration, V.~Khachatryan et~al., {\it {Measurement of spin
  correlations in t-tbar production using the matrix element method in the
  muon+jets final state in pp collisions at sqrt(s) = 8 TeV}},
  \href{http://xxx.lanl.gov/abs/1511.0617}{{\tt arXiv:1511.0617}}.

\bibitem{Andersen:2014efa}
J.~R. Andersen et~al., {\it {Les Houches 2013: Physics at TeV Colliders:
  Standard Model Working Group Report}},
  \href{http://xxx.lanl.gov/abs/1405.1067}{{\tt arXiv:1405.1067}}.

\bibitem{Bern:1994cg}
Z.~Bern, L.~J. Dixon, D.~C. Dunbar, and D.~A. Kosower, {\it {Fusing gauge
  theory tree amplitudes into loop amplitudes}},  {\em Nucl.Phys.} {\bf B435}
  (1995) 59--101, [\href{http://xxx.lanl.gov/abs/hep-ph/9409265}{{\tt
  hep-ph/9409265}}].

\bibitem{Bern:1994zx}
Z.~Bern, L.~J. Dixon, D.~C. Dunbar, and D.~A. Kosower, {\it {One loop n point
  gauge theory amplitudes, unitarity and collinear limits}},  {\em Nucl.Phys.}
  {\bf B425} (1994) 217--260,
  [\href{http://xxx.lanl.gov/abs/hep-ph/9403226}{{\tt hep-ph/9403226}}].

\bibitem{Ossola:2006us}
G.~Ossola, C.~G. Papadopoulos, and R.~Pittau, {\it {Reducing full one-loop
  amplitudes to scalar integrals at the integrand level}},  {\em Nucl.Phys.}
  {\bf B763} (2007) 147--169,
  [\href{http://xxx.lanl.gov/abs/hep-ph/0609007}{{\tt hep-ph/0609007}}].

\bibitem{Ossola:2008xq}
G.~Ossola, C.~G. Papadopoulos, and R.~Pittau, {\it {On the Rational Terms of
  the one-loop amplitudes}},  {\em JHEP} {\bf 0805} (2008) 004,
  [\href{http://xxx.lanl.gov/abs/0802.1876}{{\tt arXiv:0802.1876}}].

\bibitem{AlcarazMaestre:2012vp}
{\bf SM AND NLO MULTILEG and SM MC Working Groups} Collaboration,
  J.~Alcaraz~Maestre et~al., {\it {The SM and NLO Multileg and SM MC Working
  Groups: Summary Report}},  \href{http://xxx.lanl.gov/abs/1203.6803}{{\tt
  arXiv:1203.6803}}.

\bibitem{Ellis:2011cr}
R.~K. Ellis, Z.~Kunszt, K.~Melnikov, and G.~Zanderighi, {\it {One-loop
  calculations in quantum field theory: from Feynman diagrams to unitarity
  cuts}},  {\em Phys.Rept.} {\bf 518} (2012) 141--250,
  [\href{http://xxx.lanl.gov/abs/1105.4319}{{\tt arXiv:1105.4319}}].

\bibitem{Gluza:2010ws}
J.~Gluza, K.~Kajda, and D.~A. Kosower, {\it {Towards a Basis for Planar
  Two-Loop Integrals}},  {\em Phys.Rev.} {\bf D83} (2011) 045012,
  [\href{http://xxx.lanl.gov/abs/1009.0472}{{\tt arXiv:1009.0472}}].

\bibitem{Kosower:2011ty}
D.~A. Kosower and K.~J. Larsen, {\it {Maximal Unitarity at Two Loops}},  {\em
  Phys.Rev.} {\bf D85} (2012) 045017,
  [\href{http://xxx.lanl.gov/abs/1108.1180}{{\tt arXiv:1108.1180}}].

\bibitem{CaronHuot:2012ab}
S.~Caron-Huot and K.~J. Larsen, {\it {Uniqueness of two-loop master contours}},
   {\em JHEP} {\bf 1210} (2012) 026,
  [\href{http://xxx.lanl.gov/abs/1205.0801}{{\tt arXiv:1205.0801}}].

\bibitem{Mastrolia:2011pr}
P.~Mastrolia and G.~Ossola, {\it {On the Integrand-Reduction Method for
  Two-Loop Scattering Amplitudes}},  {\em JHEP} {\bf 1111} (2011) 014,
  [\href{http://xxx.lanl.gov/abs/1107.6041}{{\tt arXiv:1107.6041}}].

\bibitem{Badger:2012dp}
S.~Badger, H.~Frellesvig, and Y.~Zhang, {\it {Hepta-Cuts of Two-Loop Scattering
  Amplitudes}},  {\em JHEP} {\bf 1204} (2012) 055,
  [\href{http://xxx.lanl.gov/abs/1202.2019}{{\tt arXiv:1202.2019}}].

\bibitem{Badger:2013gxa}
S.~Badger, H.~Frellesvig, and Y.~Zhang, {\it {A Two-Loop Five-Gluon Helicity
  Amplitude in QCD}},  {\em JHEP} {\bf 1312} (2013) 045,
  [\href{http://xxx.lanl.gov/abs/1310.1051}{{\tt arXiv:1310.1051}}].

\bibitem{Papadopoulos:2013hra}
C.~Papadopoulos, R.~Kleiss, and I.~Malamos, {\it {Reduction at the integrand
  level beyond NLO}},  {\em PoS} {\bf Corfu2012} (2013) 019.

\bibitem{'tHooft:1978xw}
G.~'t~Hooft and M.~Veltman, {\it {Scalar One Loop Integrals}},  {\em
  Nucl.Phys.} {\bf B153} (1979) 365--401.

\bibitem{Butterworth:2014efa}
J.~Butterworth, G.~Dissertori, S.~Dittmaier, D.~de~Florian, N.~Glover, et~al.,
  {\it {Les Houches 2013: Physics at TeV Colliders: Standard Model Working
  Group Report}},  \href{http://xxx.lanl.gov/abs/1405.1067}{{\tt
  arXiv:1405.1067}}.

\bibitem{Goncharov:1998kja}
A.~B. Goncharov, {\it {Multiple polylogarithms, cyclotomy and modular
  complexes}},  {\em Math.Res.Lett.} {\bf 5} (1998) 497--516,
  [\href{http://xxx.lanl.gov/abs/1105.2076}{{\tt arXiv:1105.2076}}].

\bibitem{Remiddi:1999ew}
E.~Remiddi and J.~Vermaseren, {\it {Harmonic polylogarithms}},  {\em
  Int.J.Mod.Phys.} {\bf A15} (2000) 725--754,
  [\href{http://xxx.lanl.gov/abs/hep-ph/9905237}{{\tt hep-ph/9905237}}].

\bibitem{Goncharov:2001iea}
A.~Goncharov, {\it {Multiple polylogarithms and mixed Tate motives}},
  \href{http://xxx.lanl.gov/abs/math/0103059}{{\tt math/0103059}}.

\bibitem{Kotikov:1990kg}
A.~Kotikov, {\it {Differential equations method: New technique for massive
  Feynman diagrams calculation}},  {\em Phys.Lett.} {\bf B254} (1991) 158--164.

\bibitem{Kotikov:1991pm}
A.~Kotikov, {\it {Differential equation method: The Calculation of N point
  Feynman diagrams}},  {\em Phys.Lett.} {\bf B267} (1991) 123--127.

\bibitem{Bern:1992em}
Z.~Bern, L.~J. Dixon, and D.~A. Kosower, {\it {Dimensionally regulated one loop
  integrals}},  {\em Phys.Lett.} {\bf B302} (1993) 299--308,
  [\href{http://xxx.lanl.gov/abs/hep-ph/9212308}{{\tt hep-ph/9212308}}].

\bibitem{Remiddi:1997ny}
E.~Remiddi, {\it {Differential equations for Feynman graph amplitudes}},  {\em
  Nuovo Cim.} {\bf A110} (1997) 1435--1452,
  [\href{http://xxx.lanl.gov/abs/hep-th/9711188}{{\tt hep-th/9711188}}].

\bibitem{Gehrmann:1999as}
T.~Gehrmann and E.~Remiddi, {\it {Differential equations for two loop four
  point functions}},  {\em Nucl.Phys.} {\bf B580} (2000) 485--518,
  [\href{http://xxx.lanl.gov/abs/hep-ph/9912329}{{\tt hep-ph/9912329}}].

\bibitem{Henn:2013pwa}
J.~M. Henn, {\it {Multiloop integrals in dimensional regularization made
  simple}},  {\em Phys.Rev.Lett.} {\bf 110} (2013), no.~25 251601,
  [\href{http://xxx.lanl.gov/abs/1304.1806}{{\tt arXiv:1304.1806}}].

\bibitem{Caffo:1998du}
M.~Caffo, H.~Czyz, S.~Laporta, and E.~Remiddi, {\it {The Master differential
  equations for the two loop sunrise selfmass amplitudes}},  {\em Nuovo Cim.}
  {\bf A111} (1998) 365--389,
  [\href{http://xxx.lanl.gov/abs/hep-th/9805118}{{\tt hep-th/9805118}}].

\bibitem{Gehrmann:2000zt}
T.~Gehrmann and E.~Remiddi, {\it {Two loop master integrals for gamma* ---> 3
  jets: The Planar topologies}},  {\em Nucl. Phys.} {\bf B601} (2001) 248--286,
  [\href{http://xxx.lanl.gov/abs/hep-ph/0008287}{{\tt hep-ph/0008287}}].

\bibitem{Gehrmann:2001ck}
T.~Gehrmann and E.~Remiddi, {\it {Two loop master integrals for gamma* $\to$ 3
  jets: The Nonplanar topologies}},  {\em Nucl.Phys.} {\bf B601} (2001)
  287--317, [\href{http://xxx.lanl.gov/abs/hep-ph/0101124}{{\tt
  hep-ph/0101124}}].

\bibitem{Bonciani:2003te}
R.~Bonciani, P.~Mastrolia, and E.~Remiddi, {\it {Vertex diagrams for the QED
  form-factors at the two loop level}},  {\em Nucl. Phys.} {\bf B661} (2003)
  289--343, [\href{http://xxx.lanl.gov/abs/hep-ph/0301170}{{\tt
  hep-ph/0301170}}]. [Erratum: Nucl. Phys.B702,359(2004)].

\bibitem{Laporta:2004rb}
S.~Laporta and E.~Remiddi, {\it {Analytic treatment of the two loop equal mass
  sunrise graph}},  {\em Nucl. Phys.} {\bf B704} (2005) 349--386,
  [\href{http://xxx.lanl.gov/abs/hep-ph/0406160}{{\tt hep-ph/0406160}}].

\bibitem{Bonciani:2008az}
R.~Bonciani, A.~Ferroglia, T.~Gehrmann, D.~Maitre, and C.~Studerus, {\it
  {Two-Loop Fermionic Corrections to Heavy-Quark Pair Production: The
  Quark-Antiquark Channel}},  {\em JHEP} {\bf 07} (2008) 129,
  [\href{http://xxx.lanl.gov/abs/0806.2301}{{\tt arXiv:0806.2301}}].

\bibitem{Gehrmann:2013cxs}
T.~Gehrmann, L.~Tancredi, and E.~Weihs, {\it {Two-loop master integrals for $q
  \bar{q} \to VV$: the planar topologies}},  {\em JHEP} {\bf 1308} (2013) 070,
  [\href{http://xxx.lanl.gov/abs/1306.6344}{{\tt arXiv:1306.6344}}].

\bibitem{vonManteuffel:2013uoa}
A.~von Manteuffel and C.~Studerus, {\it {Massive planar and non-planar double
  box integrals for light Nf contributions to $ gg \to tt $}},  {\em JHEP} {\bf
  10} (2013) 037, [\href{http://xxx.lanl.gov/abs/1306.3504}{{\tt
  arXiv:1306.3504}}].

\bibitem{Henn:2014lfa}
J.~M. Henn, K.~Melnikov, and V.~A. Smirnov, {\it {Two-loop planar master
  integrals for the production of off-shell vector bosons in hadron
  collisions}},  {\em JHEP} {\bf 1405} (2014) 090,
  [\href{http://xxx.lanl.gov/abs/1402.7078}{{\tt arXiv:1402.7078}}].

\bibitem{Gehrmann:2014bfa}
T.~Gehrmann, A.~von Manteuffel, L.~Tancredi, and E.~Weihs, {\it {The two-loop
  master integrals for $q\overline{q} \to VV$}},  {\em JHEP} {\bf 1406} (2014)
  032, [\href{http://xxx.lanl.gov/abs/1404.4853}{{\tt arXiv:1404.4853}}].

\bibitem{Caola:2014lpa}
F.~Caola, J.~M. Henn, K.~Melnikov, and V.~A. Smirnov, {\it {Non-planar master
  integrals for the production of two off-shell vector bosons in collisions of
  massless partons}},  {\em JHEP} {\bf 09} (2014) 043,
  [\href{http://xxx.lanl.gov/abs/1404.5590}{{\tt arXiv:1404.5590}}].

\bibitem{Papadopoulos:2014hla}
C.~G. Papadopoulos, D.~Tommasini, and C.~Wever, {\it {Two-loop Master Integrals
  with the Simplified Differential Equations approach}},  {\em JHEP} {\bf 01}
  (2015) 072, [\href{http://xxx.lanl.gov/abs/1409.6114}{{\tt
  arXiv:1409.6114}}].

\bibitem{Gehrmann:2015ora}
T.~Gehrmann, A.~von Manteuffel, and L.~Tancredi, {\it {The two-loop helicity
  amplitudes for $ q\overline{q}^{\prime}\to {V}_1{V}_2\to 4 $ leptons}},  {\em
  JHEP} {\bf 09} (2015) 128, [\href{http://xxx.lanl.gov/abs/1503.0481}{{\tt
  arXiv:1503.0481}}].

\bibitem{Papadopoulos:2014lla}
C.~G. Papadopoulos, {\it {Simplified differential equations approach for Master
  Integrals}},  {\em JHEP} {\bf 1407} (2014) 088,
  [\href{http://xxx.lanl.gov/abs/1401.6057}{{\tt arXiv:1401.6057}}].

\bibitem{results}
available at \url{https://www.dropbox.com/s/90iiqfcazrhwtso/results.tgz?dl=0}.

\bibitem{Chetyrkin:1981qh}
K.~Chetyrkin and F.~Tkachov, {\it {Integration by Parts: The Algorithm to
  Calculate beta Functions in 4 Loops}},  {\em Nucl.Phys.} {\bf B192} (1981)
  159--204.

\bibitem{Tkachov:1981wb}
F.~Tkachov, {\it {A Theorem on Analytical Calculability of Four Loop
  Renormalization Group Functions}},  {\em Phys.Lett.} {\bf B100} (1981)
  65--68.

\bibitem{Laporta:2001dd}
S.~Laporta, {\it {High precision calculation of multiloop Feynman integrals by
  difference equations}},  {\em Int. J. Mod. Phys.} {\bf A15} (2000)
  5087--5159, [\href{http://xxx.lanl.gov/abs/hep-ph/0102033}{{\tt
  hep-ph/0102033}}].

\bibitem{Beneke:1997zp}
M.~Beneke and V.~A. Smirnov, {\it {Asymptotic expansion of Feynman integrals
  near threshold}},  {\em Nucl.Phys.} {\bf B522} (1998) 321--344,
  [\href{http://xxx.lanl.gov/abs/hep-ph/9711391}{{\tt hep-ph/9711391}}].

\bibitem{Smirnov:2002pj}
V.~A. Smirnov, {\it {Applied asymptotic expansions in momenta and masses}},
  {\em Springer Tracts Mod.Phys.} {\bf 177} (2002) 1--262.

\bibitem{Gehrmann:2015bfy}
T.~Gehrmann, J.~M. Henn, and N.~A.~L. Presti, {\it {Analytic form of the
  two-loop planar five-gluon all-plus-helicity amplitude in QCD}},
  \href{http://xxx.lanl.gov/abs/1511.0540}{{\tt arXiv:1511.0540}}.

\bibitem{Smirnov:2014hma}
A.~V. Smirnov, {\it {FIRE5: a C++ implementation of Feynman Integral
  REduction}},  {\em Comput. Phys. Commun.} {\bf 189} (2014) 182--191,
  [\href{http://xxx.lanl.gov/abs/1408.2372}{{\tt arXiv:1408.2372}}].

\bibitem{Henn:2014qga}
J.~M. Henn, {\it {Lectures on differential equations for Feynman integrals}},
  {\em J. Phys.} {\bf A48} (2015) 153001,
  [\href{http://xxx.lanl.gov/abs/1412.2296}{{\tt arXiv:1412.2296}}].

\bibitem{Lee:2014ioa}
R.~N. Lee, {\it {Reducing differential equations for multiloop master
  integrals}},  {\em JHEP} {\bf 04} (2015) 108,
  [\href{http://xxx.lanl.gov/abs/1411.0911}{{\tt arXiv:1411.0911}}].

\bibitem{BarkatouPfluegel2009}
M.~Barkatou and E.Pfl{\"u}gel, {\it {On the Moser-and super-reduction
  algorithms of systems of linear differential equations and their complexity
  }},  {\em Journal of Symbolic Computation} {\bf 44} (2009) 1017--1036.

\bibitem{BarkatouPfluegel2007}
M.~Barkatou and E.Pfl{\"u}gel, {\it {Computing super-irreducible forms of
  systems of linear differential equations via Moser-reduction: a new
  approach}},  {\em Proceedings of the 2007 international symposium on Symbolic
  and algebraic computation} {\bf ACM} (2007) 1--8.

\bibitem{Moser1960}
J.Moser, {\it {The order of a singularity in Fuchs' theory}},  {\em
  Mathematische Zeitschrift} {\bf 72} (1960) 379--398.

\bibitem{Jantzen:2012mw}
B.~Jantzen, A.~V. Smirnov, and V.~A. Smirnov, {\it {Expansion by regions:
  revealing potential and Glauber regions automatically}},  {\em Eur. Phys. J.}
  {\bf C72} (2012) 2139, [\href{http://xxx.lanl.gov/abs/1206.0546}{{\tt
  arXiv:1206.0546}}].

\bibitem{Vollinga:2004sn}
J.~Vollinga and S.~Weinzierl, {\it {Numerical evaluation of multiple
  polylogarithms}},  {\em Comput.Phys.Commun.} {\bf 167} (2005) 177,
  [\href{http://xxx.lanl.gov/abs/hep-ph/0410259}{{\tt hep-ph/0410259}}].

\bibitem{Binoth:2000ps}
T.~Binoth and G.~Heinrich, {\it {An automatized algorithm to compute infrared
  divergent multiloop integrals}},  {\em Nucl.Phys.} {\bf B585} (2000)
  741--759, [\href{http://xxx.lanl.gov/abs/hep-ph/0004013}{{\tt
  hep-ph/0004013}}].

\bibitem{Heinrich:2008si}
G.~Heinrich, {\it {Sector Decomposition}},  {\em Int.J.Mod.Phys.} {\bf A23}
  (2008) 1457--1486, [\href{http://xxx.lanl.gov/abs/0803.4177}{{\tt
  arXiv:0803.4177}}].

\bibitem{Borowka:2012yc}
S.~Borowka, J.~Carter, and G.~Heinrich, {\it {Numerical Evaluation of
  Multi-Loop Integrals for Arbitrary Kinematics with SecDec 2.0}},  {\em
  Comput.Phys.Commun.} {\bf 184} (2013) 396--408,
  [\href{http://xxx.lanl.gov/abs/1204.4152}{{\tt arXiv:1204.4152}}].

\bibitem{Borowka:2013cma}
S.~Borowka and G.~Heinrich, {\it {Massive non-planar two-loop four-point
  integrals with SecDec 2.1}},  {\em Comput.Phys.Commun.} {\bf 184} (2013)
  2552--2561, [\href{http://xxx.lanl.gov/abs/1303.1157}{{\tt
  arXiv:1303.1157}}].

\bibitem{Borowka:2015mxa}
S.~Borowka, G.~Heinrich, S.~P. Jones, M.~Kerner, J.~Schlenk, and T.~Zirke, {\it
  {SecDec-3.0: numerical evaluation of multi-scale integrals beyond one loop}},
   {\em Comput. Phys. Commun.} {\bf 196} (2015) 470--491,
  [\href{http://xxx.lanl.gov/abs/1502.0659}{{\tt arXiv:1502.0659}}].

\bibitem{Panzer:2014caa}
E.~Panzer, {\it {Algorithms for the symbolic integration of hyperlogarithms
  with applications to Feynman integrals}},  {\em Comput. Phys. Commun.} {\bf
  188} (2014) 148--166, [\href{http://xxx.lanl.gov/abs/1403.3385}{{\tt
  arXiv:1403.3385}}].

\bibitem{Baikov:1996iu}
P.~A. Baikov, {\it {Explicit solutions of the multiloop integral recurrence
  relations and its application}},  {\em Nucl. Instrum. Meth.} {\bf A389}
  (1997) 347--349, [\href{http://xxx.lanl.gov/abs/hep-ph/9611449}{{\tt
  hep-ph/9611449}}].

\bibitem{Smirnov:2003kc}
V.~A. Smirnov and M.~Steinhauser, {\it {Solving recurrence relations for
  multiloop Feynman integrals}},  {\em Nucl. Phys.} {\bf B672} (2003) 199--221,
  [\href{http://xxx.lanl.gov/abs/hep-ph/0307088}{{\tt hep-ph/0307088}}].

\bibitem{Ita:2015tya}
H.~Ita, {\it {Two-loop Integrand Decomposition into Master Integrals and
  Surface Terms}},  \href{http://xxx.lanl.gov/abs/1510.0562}{{\tt
  arXiv:1510.0562}}.

\end{thebibliography}\endgroup

\end{document}